\documentclass[12pt,letterpaper]{article}
\input{epsf}
\usepackage{graphicx,color}
\usepackage{hyperref}
\input{psfig}
\usepackage{amsmath,amssymb}
\usepackage{fancyhdr}
\usepackage{verbatim}
\newcommand\ltap{\
  \raise.3ex\hbox{$<$\kern-.75em\lower1ex\hbox{$\sim$}}\ }
\newcommand\gtap{\
  \raise.3ex\hbox{$>$\kern-.75em\lower1ex\hbox{$\sim$}}\ }

\newcommand\simge{\mathrel{%
   \rlap{\raise 0.511ex \hbox{$>$}}{\lower 0.511ex \hbox{$\sim$}}}}
\newcommand\simle{\mathrel{
   \rlap{\raise 0.511ex \hbox{$<$}}{\lower 0.511ex \hbox{$\sim$}}}}

\newcommand{\slashchar}[1]%
        {\kern .25em\raise.18ex\hbox{$/$}\kern-.70em #1}
\def\lsim{\mathrel{\raise.3ex\hbox{$<$\kern-.75em\lower1ex\hbox{$\sim$}}}}
\def\gsim{\mathrel{\raise.3ex\hbox{$>$\kern-.75em\lower1ex\hbox{$\sim$}}}}

\newcommand\CH{{\cal H}}

\newcommand\CL{{\cal L}}
\newcommand\CM{{\cal M}}

\newcommand\CO{{\cal O}}

\newcommand\CZ{{\cal Z}}

\newcommand\ul{\underline}
\newcommand\be{\begin{equation}}
\newcommand\ee{\end{equation}}
\newcommand\bea{\begin{eqnarray}}
\newcommand\eea{\end{eqnarray}}
\newcommand\ba{\begin{array}}
\newcommand\ea{\end{array}}
\newcommand\nn{\nonumber}

\newcommand{\thalf}{\textstyle{\frac{1}{2}}}

\newcommand{\tthird}{\textstyle{\frac{1}{3}}}
\newcommand{\tfourth}{\textstyle{\frac{1}{4}}}

\newcommand\dagg{\dagger}

\newcommand\gev{{\rm GeV}}
\newcommand\tev{{\rm TeV}}

\newcommand\pb{{\rm pb}}

\newcommand\fb{{\rm fb}}

\newcommand\cbeta{c_\beta}
\newcommand\sbeta{s_\beta}
\newcommand\cdelta{c_\delta}
\newcommand\sdelta{s_\delta}
\newcommand\cbetap{c_{\beta'}}
\newcommand\sbetap{s_{\beta'}}

\newcommand{\LGW}{\Lambda_{\rm GW}}

\begin{document}

\title{
\vskip -15mm
 {\Large{\bf Natural Stabilization of the\\ Higgs Boson's Mass and
     Alignment}}\\ \bigskip
} \author{ {\large Kenneth Lane$^1$\thanks{lane@bu.edu}} \, and\, William
  Shepherd$^{2,3}$\thanks{shepherd@shsu.edu}\\
  {\large $^1$Department of Physics, Boston University}\\
  {\large 590 Commonwealth Avenue, Boston, Massachusetts 02215, USA}\\
  {\large $^2$PRISMA Cluster of Excellence \& Mainz Institute of Theoretical
    Physics}\\
  {\large Johannes Gutenberg-Universit\"at Mainz, 55099 Mainz, Germany}\\
  {\large $^3$Department of Physics, Sam Houston State University}\\
  {\large Huntsville, Texas 77341, USA}
} \maketitle


\begin{abstract}

  Current data from the LHC indicate that the $125\,\gev$ Higgs boson, $H$,
  is either the single Higgs of the Standard Model or, to a good
  approximation, an ``aligned Higgs''. We propose that $H$~is the
  pseudo-Goldstone dilaton of Gildener and Weinberg. Models based on their
  mechanism of scale symmetry breaking can naturally account for the Higgs
  boson's low mass and aligned couplings. We conjecture that they are the
  only way to achieve a ``Higgslike dilaton'' that is actually {\em the}
  Higgs boson. These models further imply the existence of additional Higgs
  bosons in the vicinity of 200 to about $550\,\gev$. We illustrate our
  proposal in a version of the two-Higgs-doublet model of Lee and
  Pilaftsis. Our version of this model is consistent with published precision
  electroweak and collider physics data. We describe tests to confirm, or
  exclude, this model at Run~3 of the LHC.

  \end{abstract}


\newpage

\section*{1. The Gildener-Weinberg mechanism for \\
stabilizing the Higgs mass and alignment}

The $125\,\gev$ Higgs boson $H$ discovered at the LHC in 2012 is a
puzzle~\cite{Aad:2012tfa,Chatrchyan:2012ufa}. Its known couplings to
electroweak (EW) gauge bosons ($W$, $Z$, $\gamma$), to gluons and to fermions
($\tau$, $b$ and $t$, so far) are consistent at the 10--20\% level with those
predicted for the single Higgs of the Standard Model
(SM)~\cite{Tanabashi:2018oca,ATLAS:2018doi,Sirunyan:2018hoz,Aaboud:2018urx}. But
is that all? Why is the Higgs so light --- especially in the absence of a
shred of evidence for any new physics that could explain its low mass? Is
naturalness a chimera?

If there are more Higgs bosons --- as favored in most of the new physics
proposed to account for~$H$ and a prime search topic of the ATLAS and CMS
Collaborations --- why are $H$'s known couplings so SM-like? The common and
attractive answer is that of Higgs alignment. In the context, e.g., of a
model with several Higgs doublets,
\be\label{eq:Phii}
\Phi_i = \frac{1}{\sqrt{2}}\left(\ba{c}\sqrt{2} \phi_i^+ \\ v_i + \rho_i + i
  a_i \ea\right), \quad i = 1,2,\dots,
\ee
where $v_i/\sqrt{2}$ is the vacuum expectation value (vev) of $\Phi_i$, an
aligned Higgs is one that is a {\em mass eigenstate} given by
\be\label{eq:align}
H = \sum_{i} v_i \rho_i/v
\ee
with $v = \sqrt{\sum_i v_i^2} = 246\,\gev$. Eq.~(\ref{eq:align}) has the same
form as the linear combination of $\phi^\pm_i$ and $a_i$ eaten by the $W^\pm$
and $Z$. And this $H$ has exactly SM couplings to $W$, $Z$, $\gamma$, gluons
and the quarks and leptons.

To our knowledge, the first discussion of an aligned Higgs boson appeared in
Ref.~\cite{Gunion:2002zf}. It was discussed there in the context of a
two-Higgs-doublet model (2HDM) in the ``decoupling limit'' in which all the
particles of one doublet are very much heavier than~$v$, and so decouple from
EW symmetry breaking. The physical scalar of the lighter doublet then has SM
couplings.

There have been many papers on Higgs alignment in the literature since
Ref.~\cite{Gunion:2002zf}, including others not assuming the decoupling
limit; see, e.g., Ref.~\cite{Carena:2013ooa}. However, with only a few
exceptions, see Refs.~\cite{Dev:2014yca,Dev:2015bta,Benakli:2018vqz,
  Benakli:2018ldd}, it appears that they have not addressed an important
theoretical question: is Higgs alignment natural? Is there an approximate
symmetry which protects it from large radiative corrections? As in these
references, this might seem a separate question of naturalness than the
radiative stability of the Higgs mass, $M_H$. In fact, this question was
settled long ago: a single symmetry, spontaneously broken scale invariance
with weak explicit breaking, accounts for the Higgs boson's mass and its
alignment.

In 1973, S.~Coleman and E.~Weinberg (CW)~\cite{Coleman:1973jx} considered a
classically scale-invariant theory of a dilaton scalar with an abelian gauge
interaction, massless scalar electrodynamics. They showed that one-loop
quantum corrections can fundamentally change the character of the theory by
explicitly breaking the scale invariance, giving the dilaton a mass and a vev
and, thereby, spontaneously breaking the gauge symmetry.

In 1976, E.~Gildener and S.~Weinberg (GW)~\cite{Gildener:1976ih} generalized
CW to arbitrary gauge interactions with arbitrary scalar multiplets and
fermions, using a formalism previously invented by
S.~Weinberg~\cite{Weinberg:1973ua}. Despite the generality, their motivation
was clearly in the context of what is now known as the Standard Model. They
assumed that, due to some unknown, unspecified underlying dynamics, the
scalars $\Phi_i$ in their model have no mass terms nor cubic couplings and,
so, the model is classically scale-invariant.\footnote{We follow GW in
  assuming that all gauge boson and fermion masses are due to their couplings
  to Higgs bosons.}$^{,}$\footnote{Bardeen has argued that the classical
  scale invariance of the SM Lagrangian with the Higgs mass term set to zero
  eliminates the quadratic divergences in Higgs mass
  renormalization~\cite{Bardeen:1995kv}. That appears not to be correct. In
  any case, as far as we know, no one has yet proposed a plausible dynamics
  that produces a scale-invariant SM potential or the more general
  $V_0(\Phi)$ in Eqs.~(\ref{eq:Vzero}) and~(\ref{eq:VzeroLP})
  below. Obviously, doing that would be a great advance.}  The quartic
potential of the massless scalar fields, which are real in this notation, is
(see Ref.~\cite{Gildener:1976ih} for details)
\be\label{eq:Vzero}
V_0(\Phi) = \frac{1}{24} f_{ijkl}\, \Phi_i \Phi_j \Phi_k \Phi_l,
\ee
with dimensionless quartic couplings $f_{ijkl}$.\footnote{We assume that the
  $f_{ijkl}$ satisfy positivity conditions guaranteeing that $V_0$ has only
  finite minima. Hermiticity of $V_0$ also constrains these couplings.}  A
minimum of $V_0$ may or may not spontaneously break any continuous
symmetries. If it does, it will also break the scale invariance resulting in
a massless Goldstone boson, the dilaton. A minimum of $V_0$ does occur for
the trivial vacuum, $\Phi_i = 0$ for all~$i$. At this minimum, all fields are
massless and scale invariance is realized in the Wigner mode. However, GW
supposed that $V_0$ has a {\em nontrivial} minimum on the ray
\be\label{eq:Phiray}
(\Phi_n)_i = n_i \phi, \qquad i = 1,2,\dots
\ee
where $\sum_i n_i^2 = 1$ and $\phi > 0$ is an arbitrary mass
scale.\footnote{GW later justify this assumption along with the fact that,
  when one-loop corrections are taken into account, this provides a deeper
  minimum than the trivial one.} They did this by adjusting the
renormalization scale to have a value $\Lambda_W$ so that {\em the minimum of
  the real continuous function $V_0(N)$ {\ul{is zero}} on the unit sphere
  $N_i N_i = 1$.} If this minimum is attained for a specific unit vector
$N_i = n_i$, then $V_0(\Phi)$ has this minimum value everywhere on the
ray~(\ref{eq:Phiray}):
\be\label{eq:VzeroN}
V_0(\Phi = \Phi_n \equiv n\phi) = \frac{1}{24} f_{ijkl}\, n_i n_j n_k n_l\,
\phi^4 = 0.
\ee
Obviously, for this to be a minimum,
\be\label{eq:extr}
\left.\frac{\partial V_0(\Phi)}{\partial \Phi_i}\right\vert_{\Phi_n} =
   f_{ijkl}\, n_j n_k n_l\,\phi^3 = 0,
\ee
and the matrix
\be\label{eq:mini}
P_{ij} = \thalf f_{ijkl}\, n_k n_l
\ee
must be positive semi-definite.

Now comes the punchline: The combination $\Phi_n = n\phi$ is an eigenvector
of $P$ with eigenvalue zero. It is the dilaton associated with the
ray~(\ref{eq:Phiray}), the flat direction of $V_0$'s minimum and the
spontaneous breaking of scale-invariance. GW called the Higgs boson $\Phi_n$
the ``scalon''. Massive eigenstates of $P$ are other Higgs bosons. Any other
massless scalars have to be Goldstone bosons ultimately absorbed via the
Higgs mechanism. Then, \`a la CW, one-loop quantum corrections $V_1(\Phi)$
can explicitly break the scale invariance, picking out a definite value
$\langle\phi\rangle_0 = v$ of~$\phi$ at which $V_0+V_1$ has a minimum and
giving the scalon a mass. Including quantum fluctuations about this minimum,
\be\label{eq:Phini}
(\Phi_n)_i = n_i(v + H) + H'_i = v_i + v_i H/v +H'_i,
\ee
where, with knowledge aforethought, we name the scalon~$H$. The other Higgs
bosons $H'_i$ are orthogonal to~$H$. To the extent that $V_1$ is not a large
perturbation on the masses and mixings of the other Higgs bosons of the tree
approximation, the $H_i'$ are small components of~$\Phi_n$.\footnote{This is
  the case in the model we discuss in Sec.~2.} Thus, {\ul{the scalon is an
    aligned Higgs boson.}}\footnote{Eqs.~(5.2)--(5.6) in
  Ref.~\cite{Gildener:1976ih} show that GW recognized that the scalon has the
  same couplings to gauge bosons and fermions as the Higgs boson does in a
  one-doublet model.} Furthermore, the alignment of $H$ is protected from
large renormalizations in the same way that its mass is: by perturbatively
small loop corrections to $V_0$ and its scale invariance. While the Higgs's
alignment is apparent in the model we adopt in Sec.~2 as a concrete
example~\cite{Lee:2012jn}, this fact {\em and} its protected status are not
stressed in that paper nor even recognized in any other paper on Higgs
alignment we have seen.

From now on, we identify $H$ with the 125~GeV Higgs boson discovered at the
LHC. From the one-loop potential, i.e., first-order perturbation theory, GW
obtained the following formula for $M_H$ (which we restate in the context of
known elementary particles, extra Higgs scalars, and their electroweak
interactions):
\be\label{eq:MHGWsq}
M_H^2 = \frac{1}{8\pi^2 v^2}\left(6M_W^4 + 3M_Z^4 +\sum_{\CH}M_{\CH}^4 -
  12m_t^4\right).
\ee
Here, the sum is over Higgs bosons $\CH$ other than $H$ that may
exist. Because this is first-order perturbation theory, the masses on the
right side are those determined in zeroth order but evaluated at the
scale-invariance breaking value~$v$ of~$\phi$. For $M_H = 125\,\gev$,
Eq.~(\ref{eq:MHGWsq}) implies the sum rule
\be\label{eq:MHsum}
\left(\sum_{\CH} M_{\CH}^4\right)^{1/4} = 540\,\gev.
\ee
This result was obtained in Ref.~\cite{Hashino:2015nxa} and used in
Ref.~\cite{Lee:2012jn} to constrain the masses of new scalars. It does not
appear to have received the attention it deserves. 
It applies to all extra-Higgs models based on the GW mechanism that do not
contain additional weak bosons or heavy fermions. Thus, the more Higgs
multiplets a scalon model has, the lighter they will be. So long as loop
factors suppress the higher-order corrections to Eq.~(\ref{eq:MHsum}), it
should be a good indication of the mass range of additional Higgs bosons in
this very broad class of models.

There have been a number papers on scale invariance
leading to a ``Higgslike dilaton'' before and since the 2012 discovery of the
Higgs boson, Refs.~\cite{Goldberger:2008zz,Bellazzini:2012vz,Serra:2013kga,
  Bellazzini:2014yua,Ghilencea:2016dsl,Hernandez-Leon:2017kea} to cite
several. An especially thorough discussion is contained in
Ref.~\cite{Bellazzini:2012vz}. The authors of this paper examined the
possibility that $H(125)$ ``actually corresponds to a dilaton: the Goldstone
boson of scale invariance spontaneously broken at a scale~$f$.'' Such a
dilaton ($\sigma$) has couplings to EW gauge bosons~$W,Z$ and fermions~$\psi$
induced by its coupling to the trace of the energy-momentum tensor. They are
\be\label{eq:sigcouplings}
  \left(2g M_W W_\mu^+ W^{\mu\, -} + \sqrt{g^2+g^{\prime\,2}} M_Z Z_\mu
  Z^\mu\right)\frac{\sigma v}{f} - \sum_{\psi}
  \frac{Y_\psi}{2}\bar\psi_L\psi_R \left(1+\gamma_L+\gamma_R\right)
  \frac{\sigma v}{f} + {\rm h.c.},
\ee
where $\gamma_{L,R}$ are possible anomalous dimensions. Apart from
$\gamma_{L,R}$, these couplings are the same as those of an SM Higgs, but
scaled by $v/f$.

In general, the decay constant of a Higgslike dilaton satisfies
$|v/f| \le 1$. Ref.~\cite{Bellazzini:2012vz}, written about six months after
the discovery of $H(125)$, concluded that $|v/f| \simge 0.90$ (assuming that
$\gamma_{L,R} = 0$). Obviously, the constraint on $|v/f|$ is tighter now,
possibly $|v/f| \simge 0.95$, since all measured Higgs signal strengths,
($\sigma(H) B(H \to X))/(\sigma(H) B(H\to X))_{\rm SM}$,
would be proportional to $(v/f)^2$. An important point stressed by the
authors is that $f \simeq v$ (probably $f = v$) is achieved in models in
which {\em only} operators charged under the EW gauge group obtain vacuum
expectation values, i.e., $f = v$ only if the agent responsible for
electroweak symmetry breaking (EWSB) is also the one responsible for scale
symmetry breaking. 

A major obstacle to the Higgslike dilaton stressed in
Ref.~\cite{Bellazzini:2012vz} is that, in non-supersymmetric models it is
generally very unnatural that the dilaton's mass $M_\sigma$ is much less than
the scale $\Lambda_f \simeq 4\pi f$ of the dynamics underlying spontaneous
scale symmetry breaking.\footnote{Two exceptions are in
  Refs.~\cite{Bellazzini:2013fga,Coradeschi:2013gda}, but the models
  presented there are aimed at the cosmological constant problem and have
  nothing directly to do with EWSB.} The authors do mention that a potential
of Coleman-Weinberg type (and, by extension, Gildener-Weinberg) can naturally
achieve a large hierarchy of scales, $M_\sigma \ll 4\pi f$. But this mention
appears to be in passing because they do not provide nor cite a concrete
model that makes $f = v$ with $M_\sigma \equiv M_H \ll 4\pi v$. Nor do any of
the papers referring to Ref.~\cite{Bellazzini:2012vz}.

Interpreted in the light of Ref.~\cite{Bellazzini:2012vz}, it is clear that
that the Gildener-Weinberg mechanism is exactly a framework for obtaining a
``Higgslike dilaton'' with $f = v$ and
$M_\sigma \,\,({\rm a.k.a.\,\,} M_H) \ll 4\pi f$. We know of no other example
of this. We conjecture that the GW mechanism is the only one that can achieve
a light, aligned Higgs boson through scale symmetry breaking. It may be the
only example in which a single symmetry is responsible for both its low mass
and its alignment.

In Sec.~2 we analyze a variant of a two-Higgs-doublet model of the GW
mechanism proposed by Lee and Pilaftsis (LP) in 2012~\cite{Lee:2012jn}. In
Sec.~3 we examine constraints on our model from precision electroweak
measurements at LEP and searches for new, extra Higgs bosons at the LHC. Our
variant is consistent with all published collider data. There is much room
for improvement in those searches, and we list several targets of opportunity
both for establishing the model and for excluding it. A short Conclusion
re-emphasizes our main points. A detailed calculation of the CP-even Higgs
mass matrix and the degree to which Higgs alignment is preserved at one-loop
order, and a comparison with corresponding calculations of Lee and Pilaftsis
are reserved for an appendix.

\section*{2. The Lee-Pilaftsis model}

The Lee-Pilaftsis model employs two Higgs doublets, $\Phi_1$ and
$\Phi_2$. For reasons that will be clear in Sec.~3, we impose a type-I
$\CZ_2$ symmetry under which the scalar doublets and all SM fermions, left
and right-handed quark and lepton fields --- $\psi_L,\psi_{uR}, \psi_{dR}$
--- transform as follows:\footnote{The scalar doublets and fermion fields
  have the usual weak hypercharges $Y$ so that their electric charges are
  $Q = T_3 + Y$.}
\be\label{eq:Z2}
\Phi_1 \to -\Phi_1,\,\, \Phi_2 \to \Phi_2, \quad
\psi_L \to -\psi_L,\,\, \psi_{uR} \to \psi_{uR},\,\,
\psi_{dR} \to \psi_{dR}.
\ee
Thus, all fermions couple to $\Phi_1$ only, and there are no flavor-changing
neutral current interactions induced by Higgs exchange at tree
level~\cite{Glashow:1976nt}. Some unknown dynamics at high energies is assumed
to generate a Higgs potential that is $\CZ_2$-invariant and classically
scale-invariant, i.e., has no quadratic terms:
\bea\label{eq:VzeroLP}
V_0(\Phi_1,\Phi_2) &=&\lambda_1 (\Phi_1^\dagg \Phi_1)^2 +
\lambda_2 (\Phi_2^\dagg \Phi_2)^2 +
\lambda_3(\Phi_1^\dagg \Phi_1)(\Phi_2^\dagg \Phi_2)\nn \\
&+& \lambda_4(\Phi_1^\dagg \Phi_2)(\Phi_2^\dagg \Phi_1)+
\thalf\lambda_5\left((\Phi_1^\dagg \Phi_2)^2 + (\Phi_2^\dagg
  \Phi_1)^2\right).
\eea
All five quartic couplings are real so that $V_0$ is CP-invariant as well.

The scalars $\Phi_{1,2}$ are parameterized as in Eq.~(\ref{eq:Phii}) except
that $\Phi_{1,2}$ cannot have specific vevs $v_i$ at this stage. That
would correspond to an {\em explicit} breaking of scale invariance, and $V_0$
has no such breaking.\footnote{Here, we depart from the development in LP to
  follow the analysis in GW. We do end up in the same place as LP when the
  one-loop potential induces explicit scale symmetry breaking.} $V_0$ does
have a trivial CP and electric charge-conserving extremum at
$\Phi_1 = \Phi_2 = 0$. Following GW, we ask if there is another vacuum at
which $V_0$ vanishes, but which is nontrivial, spontaneously breaking scale
invariance. There is: consider $V_0$ on the ray
\be\label{eq:theray}
\Phi_{1\beta} = \frac{1}{\sqrt{2}} \left(\ba{c} 0\\ \phi\,\cbeta \ea\right),\quad
\Phi_{2\beta} = \frac{1}{\sqrt{2}} \left(\ba{c} 0\\ \phi\,\sbeta \ea\right).
\ee
Here $\phi > 0$ is any real mass scale, $\cbeta = \cos\beta$ and
$\sbeta = \sin\beta$, where $\beta$~is an angle to be determined. Then
\be\label{eq:Vzerobeta}
V_{0\beta} \equiv V_0(\Phi_{1\beta},\Phi_{2\beta}) =
\tfourth \left(\lambda_1 \cbeta^4  + \lambda_2 \sbeta^4
  + \lambda_{345}\cbeta^2 \sbeta^2 \right)\phi^4,
\ee
where $\lambda_{345} = \lambda_3 + \lambda_4 + \lambda_5$. We require that
$V_0$ is a minimum on this ray. The extremal (``no tadpole'') conditions are
\bea\label{eq:first}
\left.\frac{\partial V_0}{\partial \rho_1}\right\vert_{\Phi_i=
  \Phi_{i\beta}} &=& \phi^3 \cbeta\left(\lambda_1\cbeta^2 +
  \thalf\lambda_{345}\sbeta^2\right) = 0,\nn\\
\left.\frac{\partial V_0}{\partial \rho_2}\right\vert_{\Phi_i=
  \Phi_{i\beta}} &=& \phi^3 \sbeta\left(\lambda_2\sbeta^2 +
  \thalf\lambda_{345}\cbeta^2\right) = 0.
\eea
For $\beta \neq 0,\pi/2$, these conditions imply $V_{0\beta} = 0$; i.e., the
vanishing of the potential on this ray is not a separate, {\em ab initio}
assumption. These conditions also imply
\be\label{eq:extrconds}
\lambda_1/\lambda_2 = \tan^4\beta,\quad \lambda_{345} = \pm 2\sqrt{\lambda_1
  \lambda_2}.
\ee
Vacuum stability of $V_0$ requires that $\lambda_1$ and $\lambda_2$ are
positive. We shall see that non-negative eigenvalues for the CP-even Higgs
mass matrix requires $\lambda_{345} < 0$. Thus, $\lambda_{345} = -
2\sqrt{\lambda_1\lambda_2}$ at tree level. 

The matrices of second derivatives for the neutral CP-odd, charged and
CP-even scalars, respectively, are:
\bea
\label{eq:Mmsq}
\CM^2_{H_{0^-}} &=&-\lambda_5 \phi^2 \left(\ba{cc} \sbeta^2 & -\sbeta
  \cbeta\ \\ -\sbeta \cbeta& \cbeta^2 \ea\right), \\
\label{eq:Mchsq}
\CM^2_{H^\pm} &=& -\thalf\lambda_{45}\phi^2
  \left(\ba{cc} \sbeta^2 & -\sbeta \cbeta\
\\ -\sbeta \cbeta& \cbeta^2 \ea\right), \\
\label{eq:Mpsq0}
\CM^2_{H_{0^+}} &=& \phi^2  \left(\ba{cc} 2\lambda_1\cbeta^2 &
  \lambda_{345} \sbeta \cbeta\
\\ \lambda_{345}\sbeta \cbeta& 2\lambda_2 \sbeta^2 \ea\right)
 = -\lambda_{345} \phi^2  \left(\ba{cc} \sbeta^2 & -\sbeta \cbeta\
\\ -\sbeta \cbeta& \cbeta^2 \ea\right), \hspace{0.75cm}
\eea
where $\lambda_{45} = \lambda_4 + \lambda_5$ and we used
Eqs.~(\ref{eq:first}). The eigenvectors and eigenvalues of these
matrices are (taking some liberty with the eigenvalue notation):
\bea
\label{eq:mevec}
\left(\ba{c} z \\ A\ea\right) &=& \left(\ba{cc} \cbeta & \sbeta\\
     -\sbeta & \cbeta\ea\right) \left(\ba{c} a_1 \\ a_2\ea\right),
\quad M_z^2 = 0, \,\,\, M_A^2  = -\lambda_5 \phi^2;\\
\label{eq:chevec}
\left(\ba{c} w^\pm \\ H^\pm\ea\right) &=& \left(\ba{cc} \cbeta & \sbeta\\
     -\sbeta & \cbeta\ea\right) \left(\ba{c} \phi_1^\pm \\ \phi_2^\pm\ea\right),
\quad M_{w^\pm}^2 = 0, \,\,\, M_{H^\pm}^2 = -\thalf\lambda_{45}
\phi^2;\hspace{0.75cm} \\
\label{eq:pevec}
\left(\ba{c} H \\ H'\ea\right) &=& \left(\ba{cc} \cbeta & \sbeta\\
     -\sbeta & \cbeta\ea\right) \left(\ba{c} \rho_1 \\ \rho_2\ea\right),
\quad M_H^2 = 0, \,\,\, M^2_{H'}  = -\lambda_{345} \phi^2.
\eea
Positivity of the nonzero eigenvalues requires
\be\label{eq:positiv}
\lambda_5 <0, \quad \lambda_{45} < 0, \quad \lambda_{345} < 0.
\ee
So, $V_0$ has a flat minimum $V_{0\beta}$ on the ray $\Phi{i\beta}$,
degenerate with the trivial one. The conditions~(\ref{eq:positiv}) are
consistent with the convexity conditions on $V_0$~\cite{Lee:2012jn}.

The minimum\footnote{Actually, of course, the infinity of degenerate minima.}
defined by the ray in Eq.~(\ref{eq:theray}) has spontaneously broken scale
invariance. The scalar fields, $A$, $H^\pm$ and $H'$, are massive and the
massless CP-even scalar $H = \cbeta\rho_1 + \sbeta\rho_2$ is the dilaton
associated with this breaking. It is an aligned Higgs boson, the GW
scalon. The Goldstone bosons $z$ and $w^\pm$ are, of course, the longitudinal
components of the EW gauge bosons $Z$ and $W^\pm$. The minimum $V_{0\beta}$
of $V_0$ is degenerate with the trivial one. The nontrivial one-loop
corrections to $V_0$ will have a deeper minimum than the potential at zero
fields~\cite{Gildener:1976ih}.

At this stage, it is interesting that $(H,w^+,w^-,z)$ are a degenerate
quartet at the critical, zero-mass point for electroweak symmetry
breaking. It has been suggested that, if this quartet are bound states of
fermions with a new strong interaction, being close to this critical
situation gives rise to nearly degenerate isovectors that are $\rho$-like and
$a_1$-like resonances and that decay, respectively and almost exclusively, to
pairs of longitudinally polarized EW bosons and to a longitudinal EW boson
plus the $125\,\gev$ Higgs boson; see
Refs.~\cite{Lane:2015fza,Appelquist:2015vdl, Brooijmans:2016vro} for
details. We speculate that, once the scale symmetry is explicitly broken by
quantum corrections, the massive but light Higgs and the longitudinal weak
bosons remain close enough to the critical point that the diboson resonances
likely carry this imprint of their origin. Whether these resonances are
light enough to be seen at the LHC or a successor collider, we do not know
but, of course, searches for them continue, as they
should~\cite{Cavaliere:2018zcf}.

For their 2HDM, LP calculated the one-loop effective potential $V_1$ and,
following GW, extremized it along the ray~(\ref{eq:theray}). The
extremal conditions are (see Ref.~\cite{Lee:2012jn} where the effective
one-loop potential is given in their Eqs.~(17,18)):
\bea
\label{eq:extrx}
\left.\frac{\partial(V_0 + V_1)}{\partial
    \rho_1}\right\vert_{\Phi_i=\Phi_{i\beta}} &=& v^3\cbeta
\left(\lambda_1\cbeta^2 +   \thalf\lambda_{345}\sbeta^2 + \Delta\widehat
    t_1/64\pi^2\right) = 0,\nn\\
\left.\frac{\partial (V_0 + V_1)}{\partial
    \rho_2}\right\vert_{\Phi_i=\Phi_{i\beta}} &=& v^3\sbeta
\left(\lambda_2\sbeta^2 + \thalf\lambda_{345}\cbeta^2 + \Delta\widehat
    t_2/64\pi^2\right) = 0.
\eea
For the nontrivial extremum with $\beta \neq 0,\,\pi/2$, these conditions
lead to a deeper minimum,
$V_{0\beta} + V_{1\beta} < V_{0\beta} = V_0(0) + V_1(0) = 0$, picking out a
particular value~$v$ of $\phi$. This is the vev of EW symmetry breaking,
$v = 246\,\gev$, and the vevs of $\Phi_1$, $\Phi_2$ are
\be\label{eq:tanbeta}
v_1 = v\cbeta, \,\, v_2 = v\sbeta \,\,\,{\rm with}\,\,\,  \tan\beta = v_2/v_1.
\ee
The angle $\beta$ can be chosen to be in the first quadrant so that
$v_1, v_2$ are real and non-negative~\cite{Carena:2002es}. Since $v \neq 0$
explicitly breaks scale invariance, all masses and other dimensionful
quantities are proportional to the appropriate power of~it. The one-loop
functions $\Delta\widehat t_{1,2}$ are given by
\bea\label{eq:Deltat}
\Delta\widehat t_i &=&
\frac{4}{v^4}\biggl[2M_W^4\left(3\ln{\frac{M_W^2}{\LGW^2}} - 1\right)
 + M_Z^4\left(3\ln{\frac{M_Z^2}{\LGW^2}} - 1\right)
 + M_{H'}^4\left(\ln{\frac{M_{H'}^2}{\LGW^2}} - 1\right) \nn\\
&\quad& + M_A^4 \left(\ln{\frac{M_A^2}{\LGW^2}} - 1\right)
        + 2M_{H^\pm}^4 \left(\ln{\frac{M_{H^\pm}^2}{\LGW^2}} - 1\right)
        - 12m_t^4 \left(\ln{\frac{m_t^2}{\LGW^2}} - \thalf\right)\delta_{i1}
\biggr],\hspace{1.0cm}
\eea
where $M_W^2 = \tfourth g^2 v^2 = M_Z^2 \cos^2\theta_W$,
$M_{H'}^2 = -\lambda_{345} v^2$, etc. Here, $\LGW$ is the renormalization
scale at which Gildener and Weinberg's one-loop potential has a nontrivial
stationary point (and from which Eq.~(\ref{eq:MHGWsq}) and
Eq.~(\ref{eq:MHsq}) below follow). Of course, physical quantities do not
depend upon it.

Next, LP determined the one-loop-corrected mass matrices of the scalars. For
the CP-odd and charged Higgs bosons, the corrections are just the nontrivial
one-loop extremal conditions of Eqs.~(\ref{eq:extrx}), so that these mass
matrices are still given by Eqs.~(\ref{eq:Mmsq},\ref{eq:Mchsq}), but with
$\phi = v$~\cite{Lee:2012jn}.

For the CP-even mass matrix, the explicit scale breaking $\phi = v$ gives the
scalon a mass. After using the nontrivial conditions in
Eqs.~(\ref{eq:extrx}), the mass matrix is~\cite{Lee:2012jn}\footnote{It is
  improper to set $\beta = 0$ or $\pi/2$ in Eq.~(\ref{eq:Mpsq1}) and then
conclude that $\CM^2_{H_{0^+}}$ still has one zero eigenvalue. Rather, one must
use the appropriate extremal conditions for $\beta = 0$ or~$\pi/2$ to derive
the $\CM^2$ matrices at zero and one-loop order.}
\be\label{eq:Mpsq1}
\CM^2_{H_{0^+}} = v^2  \left(\ba{cc}
(2\lambda_1 + \Delta\widehat m_{11}^2/64\pi^2)\cbeta^2 &
  (\lambda_{345} +  \Delta\widehat m_{12}^2/64\pi^2)\sbeta \cbeta \\
  (\lambda_{345} +  \Delta\widehat m_{12}^2/64\pi^2)\sbeta \cbeta &
  (2\lambda_2 +  \Delta\widehat m_{22}^2/64\pi^2)\sbeta^2 \ea\right).
\ee
Here,
\bea\label{eq:Deltam}
\Delta\widehat m^2_{ij} &=&
  \frac{8}{v^4}\biggl[2M_W^4\left(3\ln{\frac{M_W^2}{\LGW^2}} + 2\right)
         + M_Z^4\left(3\ln{\frac{M_Z^2}{\LGW^2}} + 2\right)
         + M_{H'}^4 \ln{\frac{M_{H'}^2}{\LGW^2}} \nn \\
&\quad&  + M_A^4 \ln{\frac{M_A^2}{\LGW^2}}
         + 2M_{H^\pm}^4 \ln{\frac{M_{H^\pm}^2}{\LGW^2}}
         - 12m_t^4 \left(\ln{\frac{m_t^2}{\LGW^2}} + \thalf\right)
         \delta_{i1}\delta_{j1}\biggr]. \hspace{1.0cm}
\eea
The top-quark term in $\Delta\widehat m^2_{11}$ breaks the universality of
the one-loop corrections to $\CM^2_{H_{0^+}}$ but, even if that term were
absent, the scalon would still become massive because the tree-level
relations~(\ref{eq:extrconds}) are modified by the one-loop extremal
conditions:
$\lambda_1 + \thalf \lambda_{345} \tan^2\beta = \CO({\rm one\,\,loop})$, etc.

There are simple relations between $\Delta\widehat m^2_{ij}$ and $\Delta \widehat
t_i$, namely,
\bea
\label{eq:mt1}
&& \frac{\Delta\widehat m^2_{11}}{64\pi^2} = 
   \frac{2\Delta\widehat t_1}{64\pi^2} +\frac{M_H^2}{v^2}, \\
\label{eq:mt2}
&& \frac{\Delta\widehat m^2_{12}}{64\pi^2} = 
   \frac{\Delta\widehat m^2_{22}}{64\pi^2}  = 
   \frac{2\Delta\widehat t_2}{64\pi^2}
  +\frac{M_H^2}{v^2} + \frac{3 m_t^4}{2\pi^2 v^4},
\eea
where $M_H$ is the scalon mass, given below in Eq.~(\ref{eq:MHsq}). Using
Eqs.~(\ref{eq:extrx}) again, the logs and scale-dependence disappear from
$\CM^2_{H_{0^+}}$, leaving
\bea\label{eq:Mpsq2}
\left(\CM^2_{H_{0^+}}\right)_{11} &=&
      \left[(2\lambda_1 - \lambda_{345})\sbeta^2 v^2 + M_H^2\right]\cbeta^2 \nn\\
\left(\CM^2_{H_{0^+}}\right)_{22} &=&
       \left[(2\lambda_2 - \lambda_{345})\cbeta^2 v^2 + M_H^2
            + 3 m_t^4/2\pi^2 v^2\right]\sbeta^2, \nn\\
\left(\CM^2_{H_{0^+}}\right)_{12} &=&
      \left[(\lambda_{345} - 2\lambda_2)\sbeta^2 v^2 + M_H^2
             + 3 m_t^4/2\pi^2 v^2\right]\sbeta \cbeta.
\eea
The CP-even mass-eigenstates are the scalon $H_1$ and, by convention, a
heavier $H_2$ defined by
\be\label{eq:cpeven}
\left(\ba{c} H_1 \\ H_2\ea\right) = \left(\ba{cc} \cdelta& -\sdelta\\
     \sdelta& \cdelta\ea\right) \left(\ba{c} H \\ H'\ea\right)
  = \left(\ba{cc} \cbetap& \sbetap\\
     -\sbetap& \cbetap\ea\right)\left(\ba{c} \rho_1 \\ \rho_2\ea\right),
\ee
where $\beta' = \beta-\delta$ and
\be\label{eq:tanbetap}
\tan 2\beta' = \frac{\left[(\lambda_{345} -2\lambda_2)\sbeta^2 +M_H^2/v^2 +
    3m_t^4/2\pi^2 v^4\right]\sin 2\beta}
   {\left[2(\lambda_1 - \lambda_2)\sbeta^2\cbeta^2 + (M_H^2/v^2)\cos2\beta
   - 3m_t^4\sbeta^2/2\pi^2 v^4\right]}.
\ee
It is easy to check that $\beta' = \beta$ and $M_{H_1}^2 = 0$ in the tree
approximation.\footnote{Hill~\cite{Hill:2014mqa} also considered a 2HDM with
  the scale-invariant potential in Eq.~(\ref{eq:VzeroLP}). In his treatment,
  $v_1 = v_2 = 0$ at tree level, while one-loop (CW) corrections can give
  nonzero vevs to $\Phi_1,\Phi_2$. Hill chose parameters so that $v_1 \neq 0$
  but $v_2 = 0$.  This leads to a very different outcome for Hill's model
  than the one we present here. In particular, $\Phi_2$ in his model is a
  degenerate quartet of massive ``dormant'' scalars. Requiring that the $0^+$
  scalar with $v_1 = v = 246\,\gev$ is $H(125)$, Hill found from the CW
  potential that the common mass of the degenerate quartet is
  $382\,\gev$. This is exactly what one obtains from Eq.~(\ref{eq:MHsq}),
  below, by putting $M_{H'} = M_A = M_{H^\pm}$.}

\begin{figure}[h!]
\includegraphics[width=1.0\textwidth]{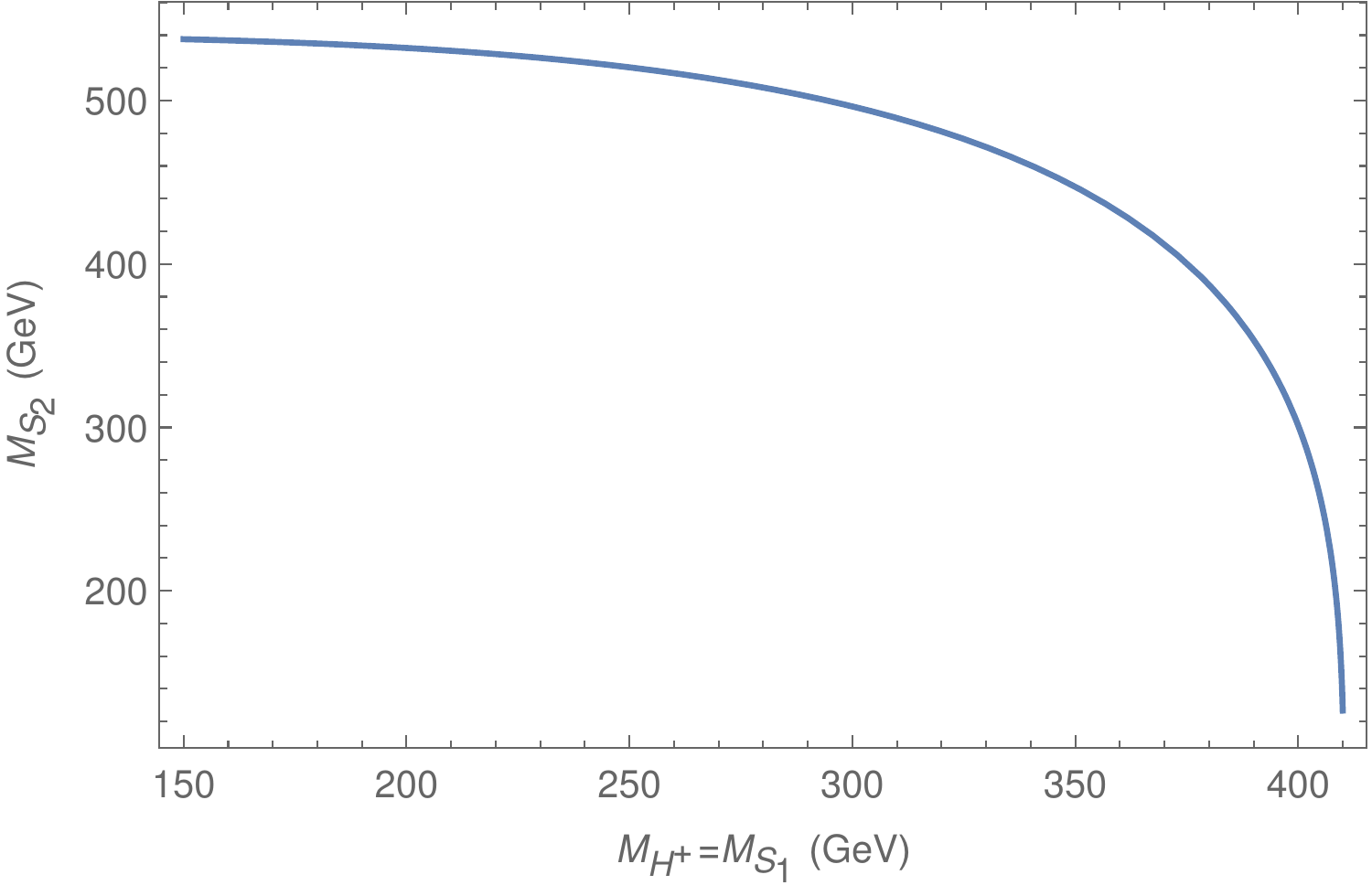}
\caption{\label{fig:mlone}The mass of the neutral Higgs $S_2 = H'/A$ as a
  function of the common mass of $H^\pm$ and the other neutral Higgs,
  $S_1 =A/H'$, from the sum rule in Eq.(\ref{eq:MHsq}) with
  $M_H = 125\,\gev$. Note the considerable sensitivity of $M_{S_2}$ to
  $M_{H^+} = M_{S_1}$ when the latter is large.}
\end{figure}

The validity of first-order nondegenerate perturbation theory requires that
$\beta' \simeq \beta$ so that $|\delta| \ll \beta$.\footnote{Below and in
  Sec.~3, experimental constraints will require $\tan\beta \simle \thalf$, and
  this means that $\delta$ must be small.} Then $H_1 \cong H$ and its mass in
this model is (from Eq.~(\ref{eq:MHGWsq}))
\be\label{eq:MHsq}
M_{H_1}^2 \cong M_H^2 = \frac{1}{8\pi^2 v^2}\left(6M_W^4 + 3M_Z^4
  + M_{H'}^4 + M_A^4 + 2M_{H^\pm}^4 - 12m_t^4\right),
\ee
where, again, all the masses on the right side of this formula are obtained
from zeroth-order perturbation theory, i.e., from $V_0$ plus gauge and Yukawa
interactions, with $\phi = v$. The way this formula is used to estimate heavy
Higgs masses is to fix the left side at $M_{H_1} = 125\,\gev$, thereby
determining $(M_{H'}^4 + M_A^4 + 2M_{H^\pm}^4)^{1/4}$. Then, as an example,
one might fix $M_{H^\pm} = M_A$ and search for $H_2 \cong H'$ {\em near} the
mass $M_{H'}$ determined by the formula. The sum rule is illustrated in
Fig.~\ref{fig:mlone} for $M_{H_1} = 125\,\gev$ and $M_{H^\pm} = M_A$ or
$M_{H'}$; the mass of the other neutral scalar is plotted
against~$M_{H^\pm}$. The figure shows that the mass of that scalar is very
sensitive to small changes in $M_{H^\pm}$ when the latter is large. In the
Appendix we compute $M_{H_1,H_2}$ as a function of $\lambda_3$, equivalently
$M^2_{H'}$, for $M_{H^\pm} = M_A = 400\,\gev$. We shall see then that there
can be appreciable differences between $M_{H'}$ and the mass eigenvalue
$M_{H_2}$ even though $M_{H_1} \cong M_H$ and the angle $\delta \ll \beta$.
Thus, the sum rule should be used with some caution in designing searches for
large values of $M_{H^\pm} = M_{A/H'}$. For this case with $M_{H'}$ diving to
zero for large $M_{H^\pm} = M_A$, using the eigenvalue $M_{H_2}$ of
$M_{H_{0^+}}$ seems the more reasonable way to estimate its mass; see
Fig.~\ref{fig:oneloop}, e.g.

The diagonalization of $\CM_{H_{0^+}}$ and the comparison of our results with
those of Ref.~\cite{Lee:2012jn} are in the Appendix. Here we mention that we
find $\delta$ and $\delta/\beta = \CO(10^{-2})$ for
$\tan\beta \simeq 1/3$--1.0, hence near perfect alignment, as we see next in
the Higgs couplings to EW bosons and fermions.

With weak hypercharges of $\thalf$, the EW gauge couplings of the physical Higgs
bosons $H_1 \cong H(125)$, $H_2$, $A$ and $H^\pm$ are given~by:
\bea\label{eq:PhiEW}
\CL_{EW} &=&  ie H^- \overleftrightarrow{\partial_\mu} H^+
              \left(A^\mu + Z^\mu \cot 2\theta_W \right) \nn\\
&+& \frac{e}{\sin 2\theta_W} A\overleftrightarrow{\partial_\mu}
    \left(H_1 \sin\delta - H_2\cos\delta\right) Z^\mu \nn\\
&+&\frac{ie}{2\sin\theta_W} \left(H^- \overleftrightarrow{\partial_\mu}
(H_1\sin\delta - H_2\cos\delta + iA)W^{+,\mu} - {\rm h.c.}\right) \nn\\
&+& (H_1\cos\delta + H_2 \sin\delta)
 \left(\frac{eM_W}{\sin\theta_W} W^{+,\mu} W^{-}_\mu +
       \frac{eM_Z}{\sin 2\theta_W} Z^\mu Z_\mu\right).
\eea
The alignment of $H_1$ and anti-alignment of $H_2$ for small~$\delta$ are
obvious.

The Yukawa couplings to mass eigenstate quarks and leptons of the physical
Higgs bosons dictated by the $\CZ_2$ symmetry in Eq.~(\ref{eq:Z2}) are given
by:
\bea\label{eq:yukawa}
\CL_Y &=& \frac{\sqrt{2}\tan\beta}{v}
      \sum_{k,l=1}^3\left[H^+\left(\bar u_{kL} V_{kl}\,m_{d_l}d_{lR}
      -\bar u_{kR}\, m_{u_k} V_{kl}\, d_{lL} +
      m_{\ell_k}\bar\nu_{kL}\ell_{kR}\,\delta_{kl}\right) + {\rm
      h.c.}\right] \nn\\
   &-& \left(\frac{v\cos\beta + H_1\cos\beta'-
       H_2\sin\beta'}{v\cos\beta}\right)
       \sum_{k=1}^3 \left(m_{u_k} \bar u_k u_k + m_{d_k} \bar d_k d_k
                         +m_{\ell_k}\bar\ell_k \ell_k\right) \nn\\
   &-& \frac{i A\tan\beta}{v} \sum_{k=1}^3 \left(m_{u_k} \bar u_k \gamma_5 u_k
     - m_{d_k} \bar d_k\gamma_5 d_k - m_{\ell_k}\bar\ell_k
     \gamma_5\ell_k\right).
\eea
Here, $V$ is the Cabibbo-Kobayashi-Maskawa matrix and fermion masses are
to be evaluated at $\CO(300\,\gev)$. Again the alignment of $H_1$ is obvious
for small~$\delta$.

The charged Higgs couplings in Eq.~(\ref{eq:yukawa}) contribute to
$b\to s\gamma$ decays. Ref.~\cite{Misiak:2006zs} studied this transition and
bounded $M_{H^\pm} > 295\,\gev$ at the 95\%~CL in 2HDM with type-II
couplings, i.e., in which up-quarks get their mass from $\Phi_2$ and
down-quarks from $\Phi_1$~\cite{Branco:2011iw}. Their bound is for
$\tan\beta \ge 2$ in such a model. The Yukawa couplings of our model are the
variant of type-I with $\Phi_1$ and $\Phi_2$ interchanged. The bound then
corresponds to $\tan\beta \le \thalf$. In Sec.~3.2 we find a similar bound on
$\tan\beta$ from a search for $H^\pm$.

We briefly mention two theoretical constraints on this model
considered in Ref.~\cite{Lee:2012jn}. The first is perturbative
unitarity. One of its most stringent conditions comes from requiring that the
eigenvalue $a_+$ of the scattering amplitudes in Ref.~\cite{Kanemura:1993hm}
obeys the bound
\be\label{eq:aplus}
a_+ = \frac{1}{16\pi}\left[3(\lambda_1 + \lambda_2) +
      \sqrt{9(\lambda_1-\lambda_2)^2+(2\lambda_3+\lambda_4)^2}\right] \le \thalf.
\ee
Note that this is symmetric under $\cbeta \leftrightarrow \sbeta$. Assuming,
e.g, that $M_{H^\pm} = M_A = 400\,\gev$, we have
\be\label{apvalues}
a_+ = \left\{\begin{array}{l} 0.38 \,\,{\rm for}\,\, \tan\beta = \thalf\\
                              0.82 \,\,{\rm for}\,\, \tan\beta =
               \tthird\end{array}\right.
\ee

The second constraint comes from the oblique parameters
$S,T$~\cite{Kennedy:1988sn,Peskin:1990zt,Peskin:1991sw,Golden:1990ig,
  Holdom:1990tc,Altarelli:1991fk}. We note here that the contribution to~$T$
from the Higgs scalars in this model vanishes identically when
$\lambda_4 = \lambda_5$~\cite{Battye:2011jj, Pilaftsis:2011ed}. For this
reason, we often assume $M_{H^\pm} = M_A$ in the phenomenological
considerations of Sec.~3. The constraints following from the $S$-parameter
will be discussed there as well.

\section*{3. Experimental constraints and opportunities}

In this section, we discuss constraints from precision EW measurements at LEP
and searches for new charged and neutral Higgs bosons at the LHC and,
finally, we summarize targets of opportunity at the LHC.

\subsection*{3.1 Precision Electroweak Constraints}

The constraints from $Z$ and $W$ boson properties~\cite{Patrignani:2016xqp},
parameterized by $S$ and $T$ are independent of the choice of Yukawa
couplings for the 2HDM. We follow Ref.~\cite{Lee:2012jn} to evaluate the
contributions of the new Higgses to these parameters which included the
(formally) two-loop effect of vertex corrections which arise due to the
potentially large quartic couplings. The general form of these corrections
is~\cite{Toussaint:1978zm}
\begin{eqnarray}
  \label{eq:STphi}
S_\Phi \!&=&\! -\frac{1}{4\pi} \left[
\left(1+\delta_{\gamma Z}^{H^\pm}\right)^2F^\prime_\Delta(M_{H^\pm},M_{H^\pm})
-\sum_{i=1,2}\left(g_{H_iAZ}+\delta_Z^{H_i}\right)^2 F^\prime_\Delta(M_{H_i},M_A)
\right]\,, \nonumber\\
T_\Phi \!&=&\! -\frac{\sqrt{2}G_F}{16\pi^2\alpha_{\rm EM}}\ \Bigg\{
-\left(1+\delta_W^{A}\right)^2F_\Delta(M_A,M_{H^\pm})\\
&&
+\sum_{i=1,2}\left[
\left(g_{H_iAZ}+\delta_Z^{H_i}\right)^2 F_\Delta(M_{H_i},M_A)
-\left(g_{H_iH^- W^+}+\delta_W^{H_i}\right)^2 F_\Delta(M_{H_i},M_{H^\pm})
\right] \Bigg\}\; ,\quad   \nonumber
\end{eqnarray}
where $\delta_V^H$ is the vertex correction to the coupling of the vector
boson $V$ to Higgs boson~$H$ (see Ref.~\cite{Lee:2012jn}) and
$F^{\left(\prime\right)}_\Delta\left(M_1,M_2\right)$ are the bubble-graph
integrals given in Ref.~\cite{Kanemura:2011sj}. As noted, the Higgs
contribution to~$T$ vanishes in this model when $M_{H^\pm} = M_A$.

\begin{figure}[t!]
\includegraphics[width=0.50\textwidth]{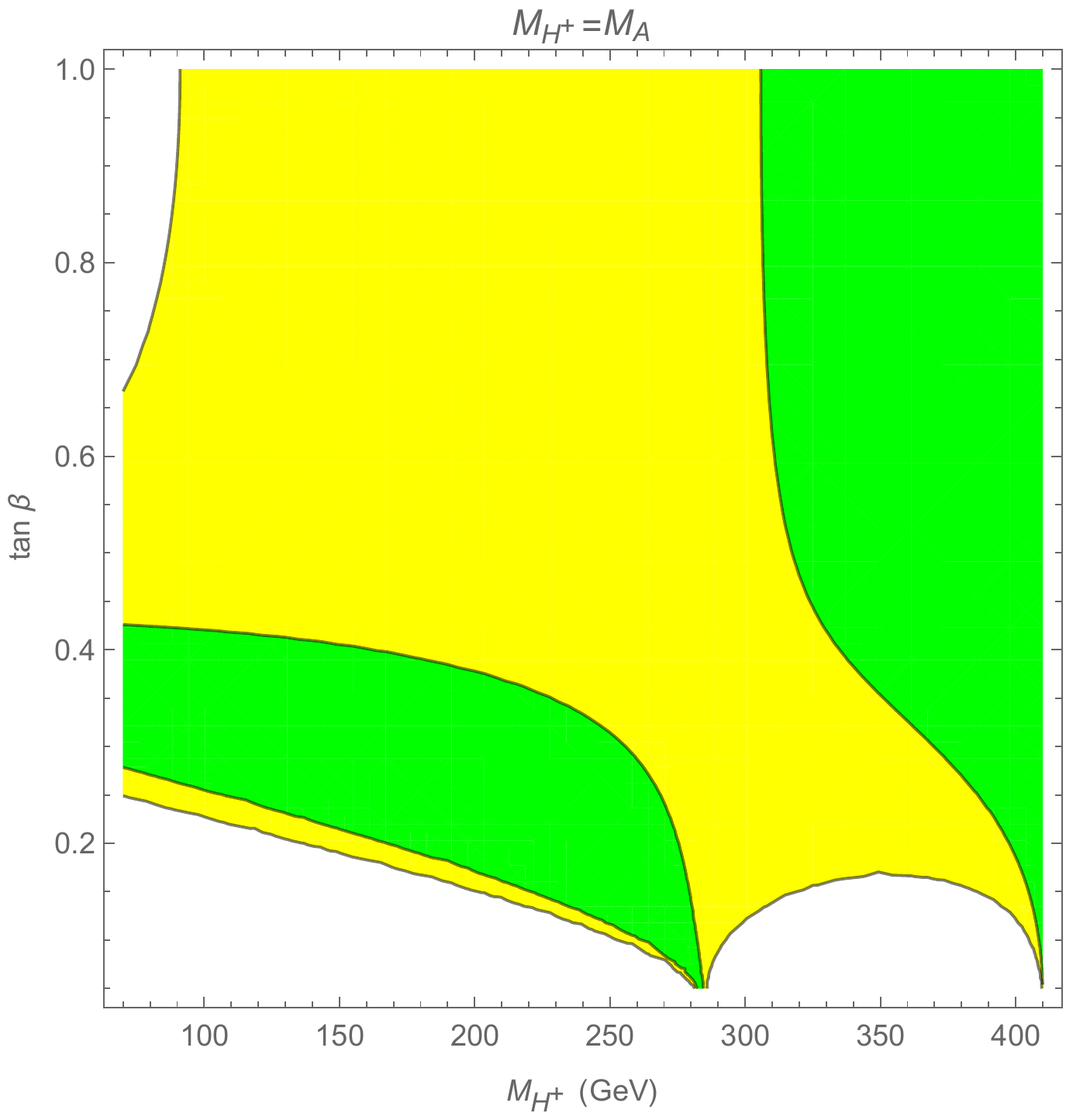}
\includegraphics[width=0.50\textwidth]{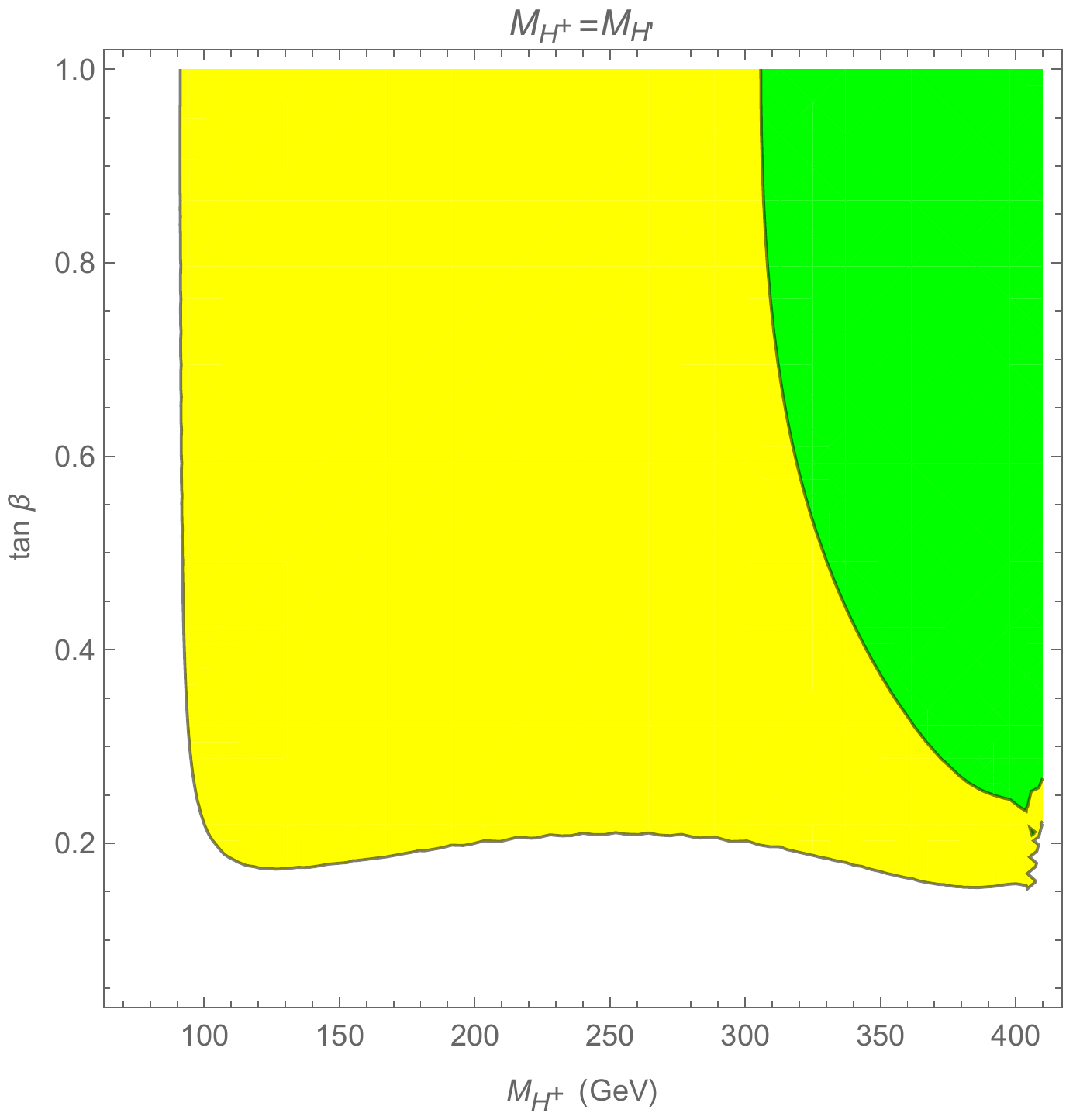}
\caption{\label{fig:PEWD} The constraints on the type-I scale-invariant 2HDM
  arising from precision electroweak measurements. The green (yellow) regions
  indicate $1\sigma$ ($2\sigma$) agreement with precision data. In the left
  panel $M_{H^+}=M_A$, which also enforces that the scalars' contribution
  $T_\Phi$ to the $T$-parameter vanishes. In the right panel
  $M_{H^+}=M_{H^\prime}$ for which $T_\Phi \neq 0$. The remaining neutral
    scalar's mass is set by Eq.~(\ref{eq:MHsq}).}
\end{figure}

The regions of $\tan\beta$--$M_{H^\pm}$ parameter space allowed by precision
EW data for the cases $M_{H^+}=M_A$ and $M_{H^+}=M_{H^\prime}$ are shown
in~Fig.~\ref{fig:PEWD}. The mass of the lone neutral scalar in either of
these scenarios is taken from the sum rule~(\ref{eq:MHsq}); see
Fig.~\ref{fig:mlone}. The axes in Fig.~\ref{fig:PEWD} are chosen to span the
parameter space technically available to the model after direct LEP
searches. The lower bound of $70\,\gev$ corresponds to the LEP search for
charged Higgses~\cite{Abbiendi:2013hk}. The upper limit of $410\,\gev$ is
chosen to avoid the region of low $M_{H'}$ or $M_A$ in Fig.~\ref{fig:mlone}.
For $M_{H^\pm} = M_{H'}$, the shared mass must be greater than about
$315\,\gev$ to satisfy EW precision data constraints at the $1\sigma$ level,
and the higher masses allow for smaller values of $\tan\beta$. In the
$M_{H^+}=M_A$ case, a similar region in shared mass and $\tan\beta$ is
allowed, but there is a second region within 1$\sigma$ at low common mass and
$\tan\beta$. In fact, nearly the entire possible mass range is allowed at the
$2\sigma$ level for $\tan\beta > 0.2$. We shall see below in
Fig.~\ref{fig:H+tb} that a CMS search at $8\,\tev$ for a charged Higgs boson
decaying to $t\bar b$ requires $\tan\beta \simle 0.5$ for
$180 < M_{H^\pm} < 500\,\gev$~\cite{Khachatryan:2015qxa}.

\subsection*{3.2 Direct Searches at the LHC}

\begin{figure}[ht!]
\includegraphics[width=1.0\textwidth]{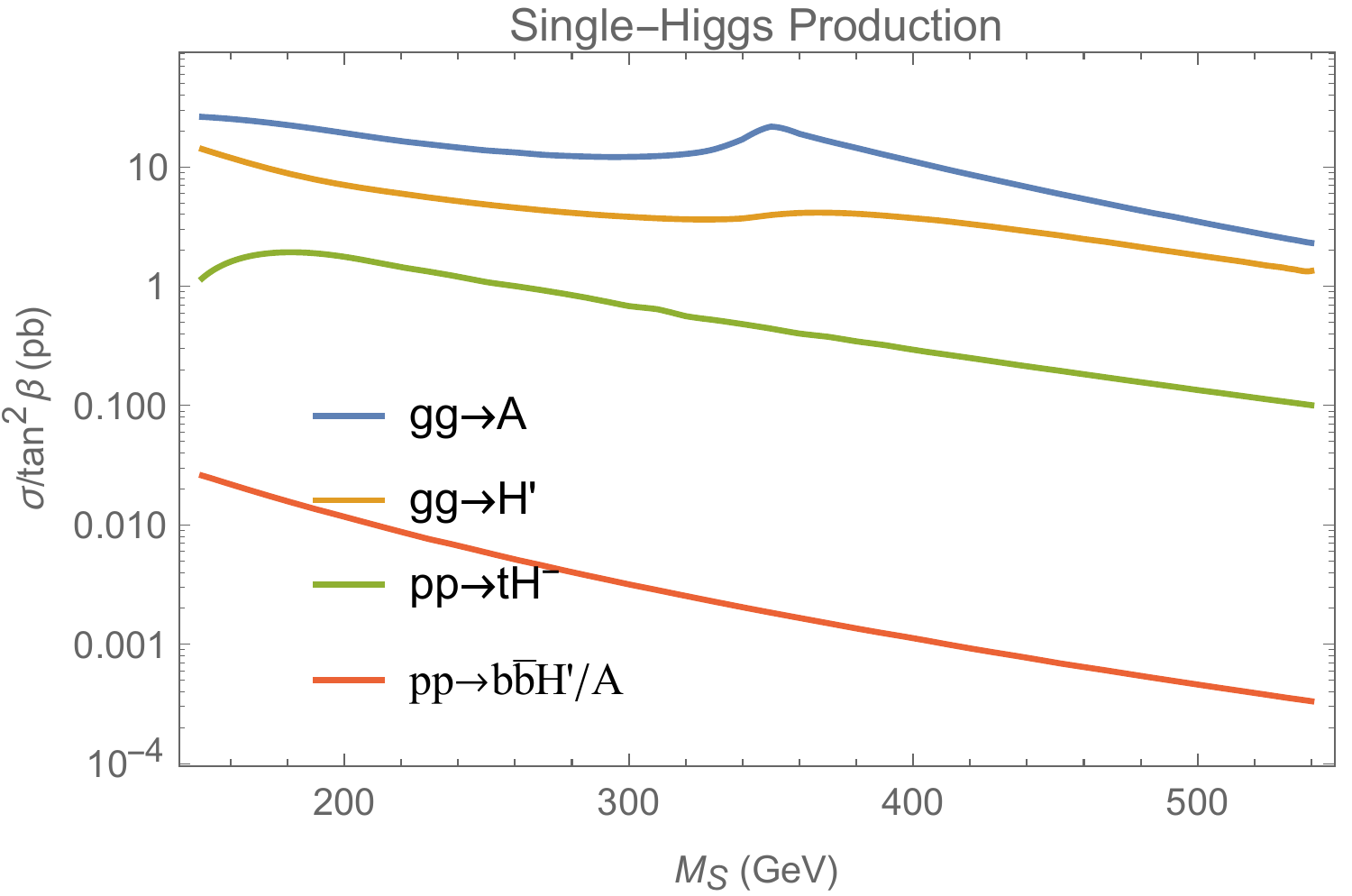}
\caption{\label{fig:sigmas}The cross sections for single Higgs production
  processes in the alignment limit ($\delta \to 0$) with the dependence on
  $\tan\beta$ scaled out. Both charged Higgs states are included in
  $pp \to t H^-$.}
\end{figure}

\begin{figure}[ht!]
\includegraphics[width=1.0\textwidth]{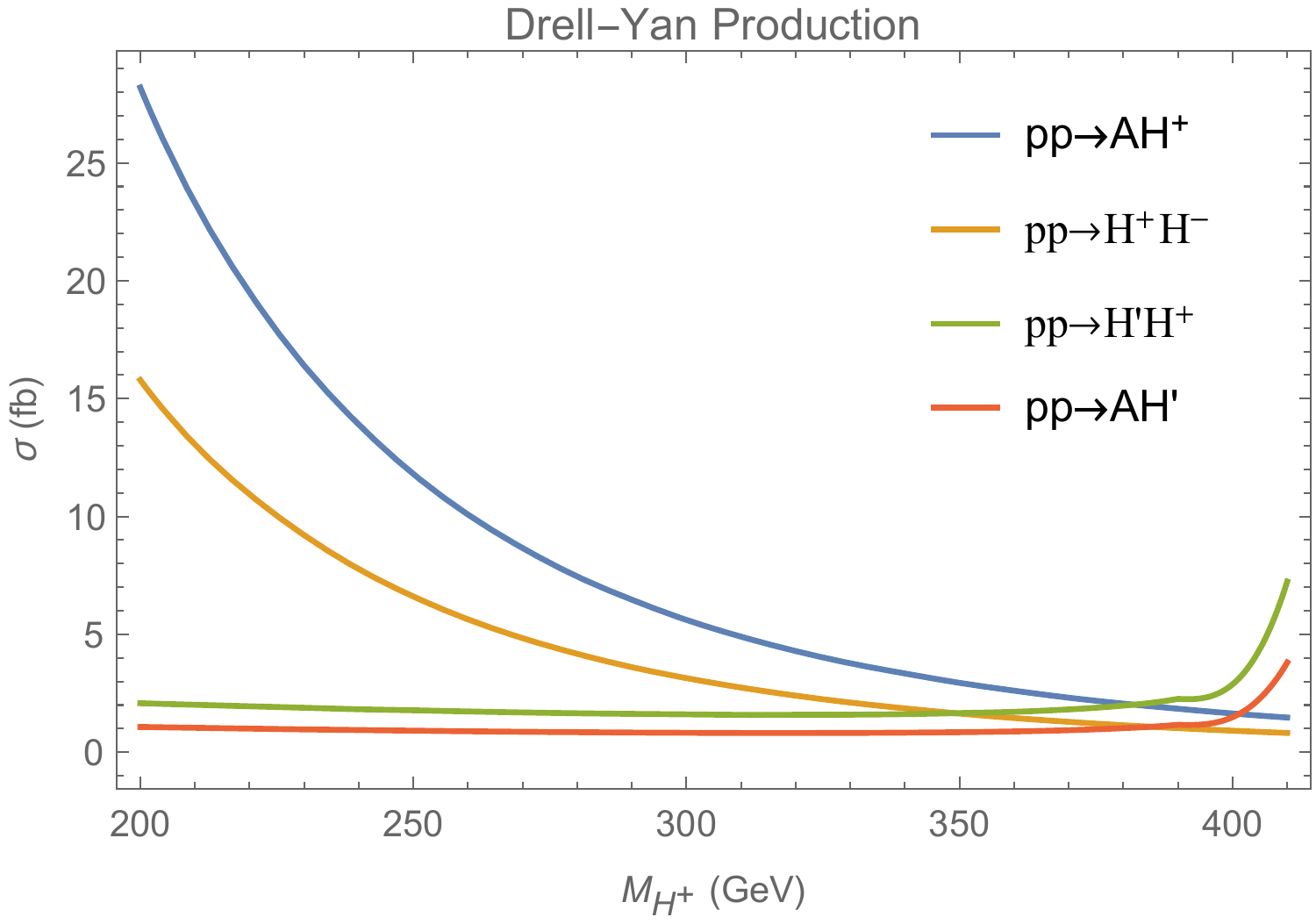}
\caption{\label{fig:DY} The cross sections for Drell-Yan production of Higgs
  pairs in the alignment limit ($\delta \to 0$). They are independent of
  $\tan\beta$.  $M_{H^\pm} = M_A$ is assumed, with $M_{H'}$ taken from
  Eq.(\ref{eq:MHsq}). The sharp increase at large $M_{H^\pm}$ is due to the
  rapid decrease of $M_{H'}$ there; see Fig.~\ref{fig:mlone}. For the case
  $M_{H^\pm} = M_{H'}$, simply interchange the labels $A$ and $H'$ in the figure.}
\end{figure}

In the alignment limit (small $\delta/\beta$), the Yukawa couplings of the
new charged and neutral Higgs bosons are proportional to $\tan\beta$. The
strong alignment renders ineffective existing searches for such Higgses in
weak boson final states, specifically $H_2 \simeq H'$ and $A \to W^+ W^-$,
$ZZ$. At the same time, it may strengthen searches in fermionic final
states. Reference production cross sections for the new Higgses for several
potentially important processes are shown in
Figs.~\ref{fig:sigmas},\ref{fig:DY}. Note that all the single-Higgs
production cross sections which may be efficient in the alignment limit are
proportional to $\tan^2\beta$.

\begin{figure}[ht!]
\includegraphics[width=1.0\textwidth]{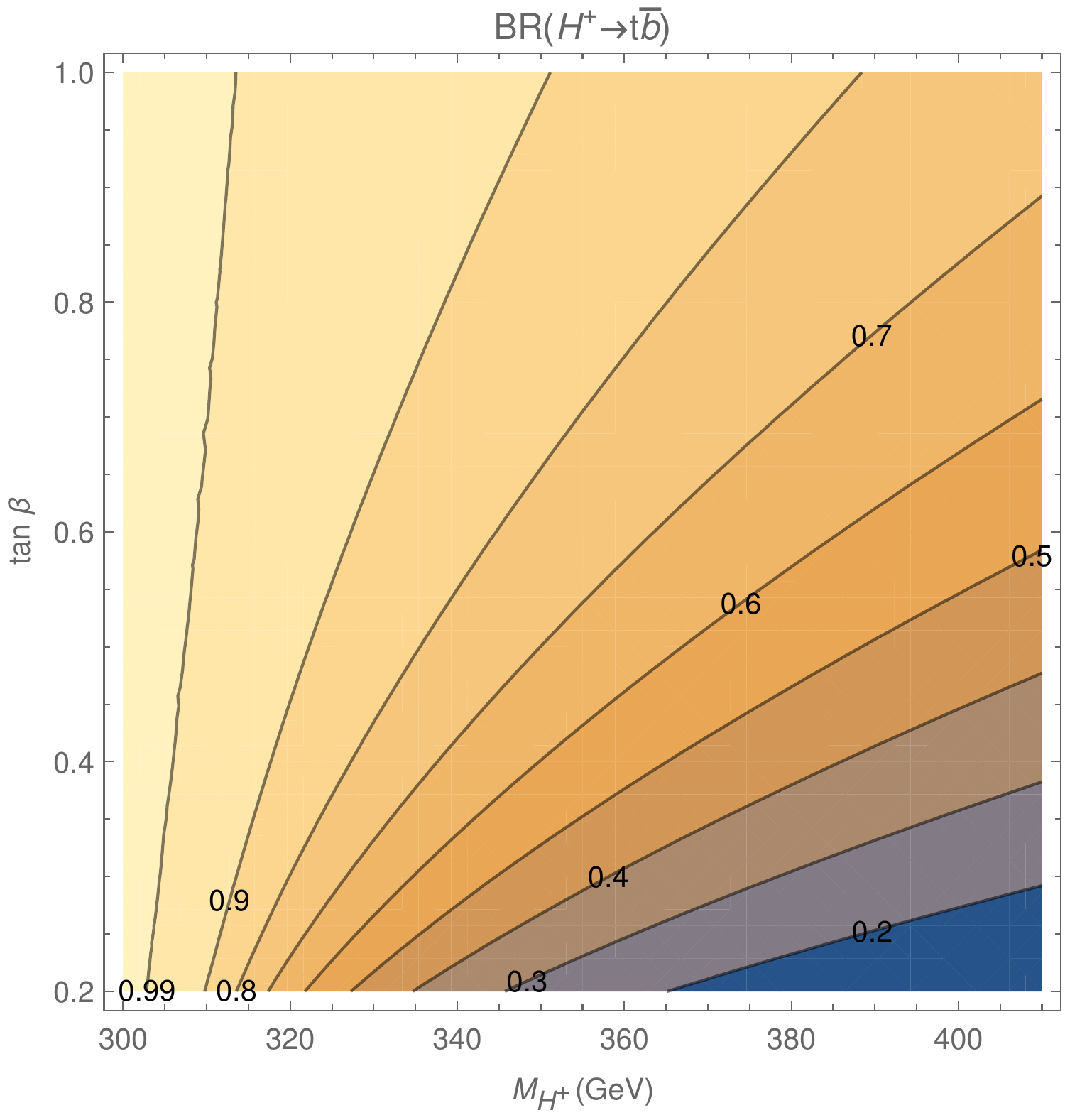}
\caption{\label{fig:BRHP}The branching ratio $B(H^+ \to t\bar b)$ as a
  function of $M_{H^+}$ and $\tan\beta$. The only other significant decay
  mode is $H^+ \to W^+ H'/A$. It is assumed that $M_{H^+} = M_A$ or $M_{H'}$;
  the other neutral scalar's mass is then given by Eq.~(\ref{eq:MHsq}).}
\end{figure}

Among heavy scalars, the most promising search is for $t H^\pm$-associated
production, with $H^+ \to t \bar b$. The subprocess for this is
$g b(\bar b) \to t H^- \, (\bar t H^+)$. The most stringent constraint so far
on this channel is from the CMS search at
$8\,\tev$~\cite{Khachatryan:2015qxa}. In the aligned limit, the other
potentially important decay mode is $H^\pm \to W^\pm H'$ or $W^\pm A$,
whichever neutral Higgs is lighter. That neutral Higgs decays mainly to
$\bar bb$.\footnote{It also decays to $\tau^+\tau^-$ with a branching ratio
  of $\CO(15\%)$.}  Fig.~\ref{fig:BRHP} shows the dependence of the
branching ratio $B(H^+ \to t\bar b)$ as a function of $M_{H^+}$ and
$\tan\beta$. In this figure, $M_{H^+} = M_A$ or $M_{H'}$ and the other
neutral scalar's mass is given by the sum rule~(\ref{eq:MHsq}). There has
been no dedicated search yet for $H^\pm \to W^\pm H'/A$. However, the final
state for this decay mode, $t H^\pm \to t W^\pm H'/A \to t W^\pm \bar bb$, is
similar to that of $t H^\pm \to tt\bar b \to t W^\pm \bar bb$. Therefore, we
{\em conservatively} assume that it contributes with equal acceptance to the
search at CMS so that the branching ratio of
$B(t H^\pm \to t W^\pm \bar bb) = 1$. The signal rate then scales as for the
single Higgs production, $\sigma\cdot B \propto\tan^2\beta$.

\begin{figure}[ht!]
\includegraphics[width=1.0\textwidth]{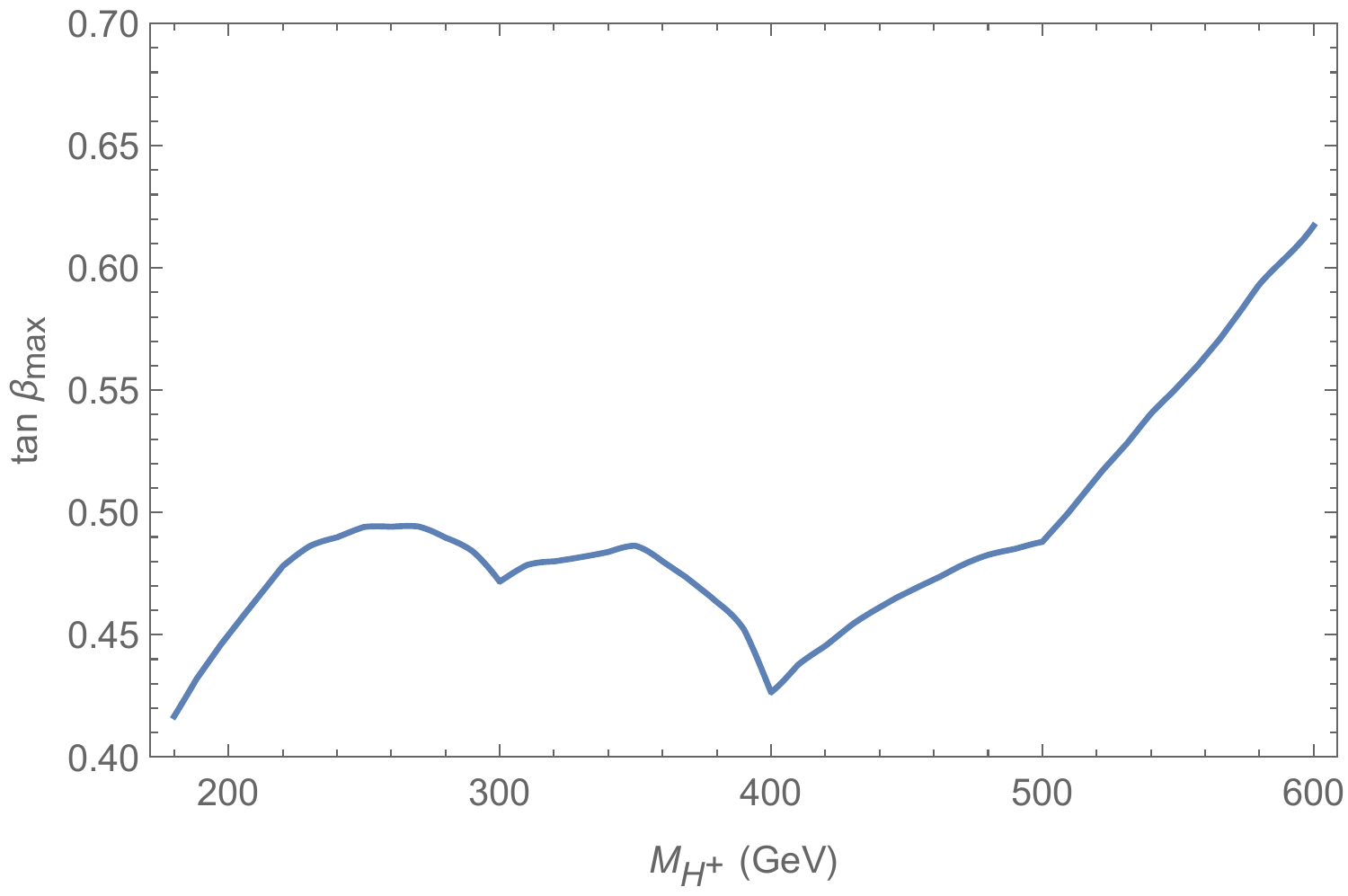}
\caption{\label{fig:H+tb}Constraints on $\tan\beta$ in any type-I 2HDM \`a la
  the model of Sec.~2 from a CMS search at $8\,\tev$ for charged Higgs
  production in association with another top quark and decaying to
  $t \bar b$~\cite{Khachatryan:2015qxa}. The kinks in this plot occur at the
  data points provided by CMS, and arise due to linear interpolation of the
  excluded cross section for intermediate values of charged Higgs mass.}
\end{figure}

Because CMS has reported unfolded bounds on $\sigma\cdot B$ for this final
state, we are able to recast the search from its type-II 2HDM form into
bounds on our type-I model. We show the constraints on $\tan\beta$ as a
function of $M_{H^+}$ in Fig.~\ref{fig:H+tb}. As did CMS, we extrapolated
linearly between points at which cross section limits were reported.

Further constraints can come from searches for neutral Higgs bosons produced
in gluon fusion and decaying to top, bottom or tau pairs.\footnote{Note that
  $WW$ and $ZZ$ fusion of $H'$ and $A$ is very small in the alignment limit.}
From the sum rule in Eq.~(\ref{eq:MHsq}), the heaviest $H'$ or $A$ can be is
almost $540\,\gev$ when all other masses are $\sim 100\,\gev$. An ATLAS
search at $8\,\tev$~\cite{Aaboud:2017hnm} for resonant production of
$\bar tt$ has been performed only for scalars heavier than $500\,\gev$
because of the complexity of interference effects in regions near threshold
where off-shell tops become important in heavy Higgs decays. A neutral scalar
mass of $M_{H'}=500\,\gev$ corresponds to $M_{H^\pm} = M_A = 295\,\gev$. In
principle, such searches are sensitive to the full mass range
within~$1\sigma$ at lower mass and $\tan\beta$ in the left panel of
Fig.~\ref{fig:PEWD}. The case $M_{H^\pm} = M_H = 295\,\gev$ is within
$2\sigma$ of Fig,\ref{fig:PEWD} and should not be ignored out of hand. This
particular search is fairly difficult to recast because the analysis was
performed primarily in terms of signal strength in a~2HDM. Using the
``signal'' rate quoted in auxiliary material together with the constrained
signal strength leads to constraints at fixed $M_A=500\,\gev$ of
$\sigma\cdot B <0.32$--$1.69\,\pb$, corresponding to
$\tan\beta < 0.62$--$1.02$ in our model. For $M_{H'} = 500\,\gev$, the limits
are $\sigma\cdot B < 0.085$--$0.40\,\pb$, corresponding to
$\tan\beta<0.59-0.91$. Choosing even the smallest of these bounds on
$\tan\beta$, this search does not reach the $1\sigma$ region at low
$M_{H^\pm} = M_A$ in Fig.~\ref{fig:PEWD}.

\begin{figure}[ht!]
\includegraphics[width=1.0\textwidth]{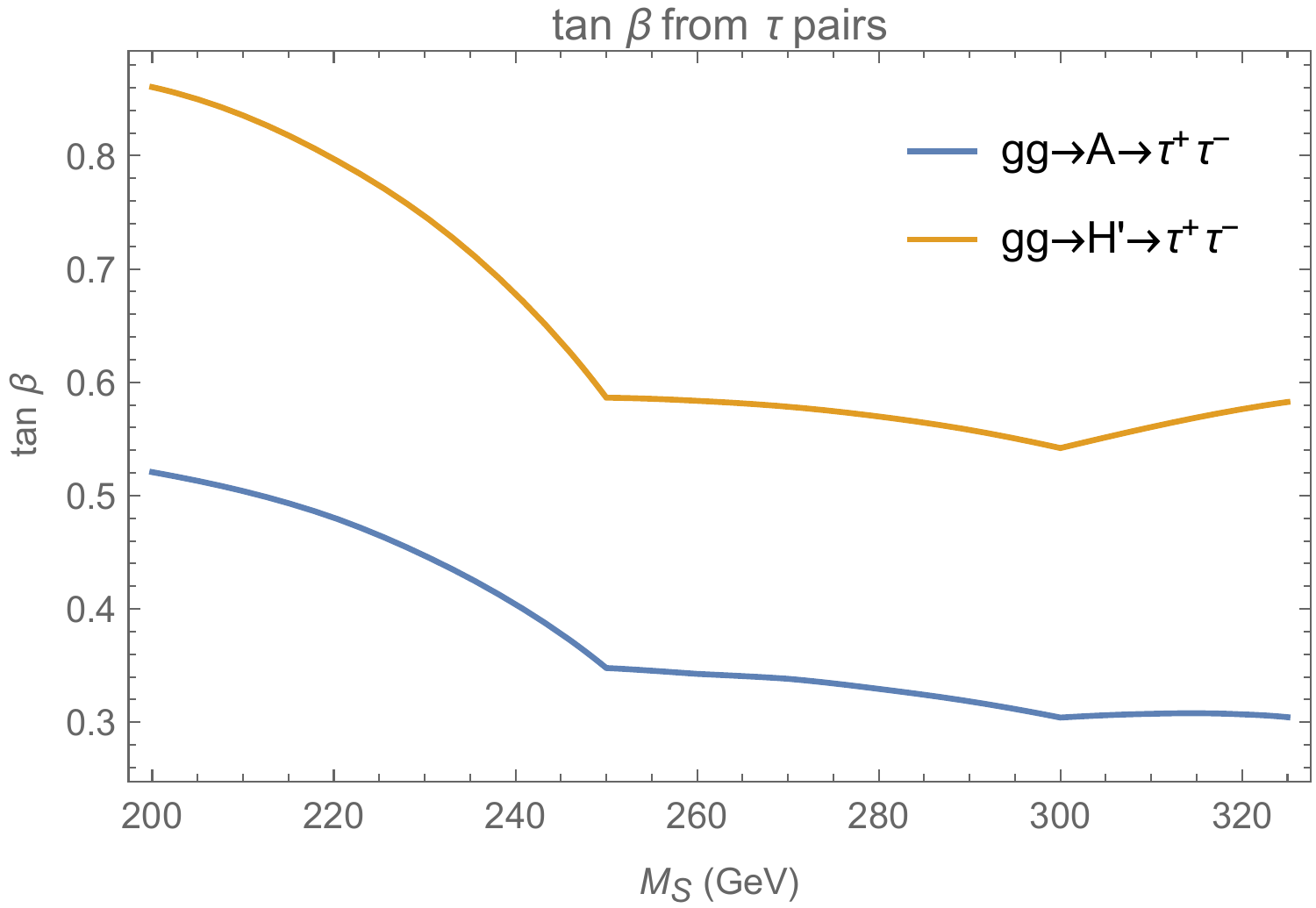}
\caption{\label{fig:tbtau}The constraints on $\tan\beta$ in the model of
  Sec.~2 from the ATLAS search at $13\,\tev$ for a neutral Higgs $S = H_2, A$
  decaying to $\tau^+\tau^-$~\cite{Aaboud:2017sjh}. They apply only to the
  lighter neutral Higgs because decays to light fermions are quickly
  overwhelmed by, e.g., $A \to H_2 Z$ or $A \to \bar tt$ when the channels
  open.}
\end{figure}

A search at $13\,\tev$ for production of a neutral scalar in association with
a $\bar bb$ pair and decaying to another $\bar bb$ pair in the mass range
$300$--$1300\,\gev$ has been carried out by CMS~\cite{Sirunyan:2018taj}. It
is not appreciably sensitive to these models, as the bottom Yukawa coupling
is not enhanced as it is in the models targeted by this analysis. The largest
$\bar bb$-associated new Higgs production cross section in our model,
independent of subsequent decay branching ratios, is already sub-femtobarn
for $\tan\beta=1$ and, so, is unconstrained by this analysis.

A search for neutral Higgs production --- from either gluon fusion or
$\bar bb$-associated production --- with subsequent decays to $\tau^+\tau^-$
has been performed at $13\,\tev$ by ATLAS in the mass range
200--$2250\,\gev$~\cite{Aaboud:2017sjh}. Gluon fusion production is more
promising for our model, with cross sections as large as $20\,\pb$ for
pseudoscalar production at $\tan\beta=1$. Decays to light fermions in this
model are quickly overwhelmed by bosonic decays, such as $A \to H_2 Z$, when
accessible. Thus, these searches are capable of constraining only the lighter
new neutral scalar in the model. In this limit, the competing decays are to
third-generation quarks. The bounds on $tan\beta$ arising from these searches
are shown in Fig.~\ref{fig:tbtau}. Due to the opening of the top quark decay
channel, these searches also become ineffective for
$M_{H_2,A} \simge 350\,\gev$.

Finally, two searches for a neutral scalar produced in association with
$\bar bb$ and decaying to $Z$ plus another scalar which itself decays to
$\bar bb$ or $\tau^+\tau^-$ has been performed at $13\,\tev$ by CMS and
ATLAS~\cite{Khachatryan:2016are,Aaboud:2018eoy} for models with both scalars'
masses below $1\,\tev$. In order that there is adequate splitting between the
scalars in our model, either the common scalar mass of the charged and
selected neutral scalar must be greater than $\sim 400\,\gev$ or less than
$\sim 350\,\gev$, implying that the heavier scalar's mass is at least
$400\,\gev$. From Fig.~\ref{fig:sigmas}, the greatest production cross
section for $pp \to b\bar b H'/A$ for $\tan\beta = 1$ is $\sim 10\,\fb$. The
CMS cross section limits (and comparable ones from ATLAS) for the lighter
scalar decaying into $\bar bb$ are greater than this largest possible cross
section. Limits for decays to $\tau^+\tau^-$ are a~few~fb. Including the tau
branching ratio of the $H'/A$, this limit is also well above the cross
section predicted in our model.

A search of interest to ATLAS and CMS is for resonant pair-production of
$H(125)$. Unfortunately, the amplitude for $H_2 \to H_1 H_1$ vanishes in the
alignment limit of 2HDM models of type considered here and, so, we expect
that it will be a very weak signal. This is related to the vanishing of
$H_2 \to WW$ and $ZZ$ in this limit. As noted in Sec.~2, before the explicit
scale-breaking potential~$V_1$ is turned on, $(H,w^+,w^-,z)$ are a degenerate
quartet at the critical zero-mass point for electroweak symmetry
breaking. Therefore, the three-point amplitude coupling of
$H' = \lim_{\delta \to 0} H_2$ to any pair of these Goldstone bosons vanishes.

\subsection*{3.3 Targets of opportunity at the LHC}

We summarize here the likely targets of opportunity at the LHC that we
discussed above and remind the reader of some unlikely ones which serve as
negative tests of the model we considered. We preface this by recalling that
we found that $\tan\beta \simle \thalf$ and this suppresses certain production
rates and decay branching ratios relative to those for the value
$\tan\beta = 1$ assumed in many 2HDM searches at the LHC.

\begin{itemize}

\item[1.)] Update the search carried out in Ref.~\cite{Khachatryan:2015qxa}
  for $H^+\to t\bar b \to W^+ \bar b b$ via the process
  $g\bar b \to \bar t H^+$ and charge conjugate
  modes.\footnote{Ref.~\cite{Aaboud:2018cwk}, which appeared recently, is a
    13~TeV search by ATLAS which addresses this final state. However, its
    bounds at low masses are not appreciably stronger than those of
    Ref.~\cite{Khachatryan:2015qxa}.}

\item[2.)] Perform a dedicated search for $g \bar b \to \bar t H^+$ followed
  by $H^+\to W^+H_2/A \to W^+b\bar{b}$. Recall that this has a similar
  final state as the search above, but includes a resonant $\bar bb$ signal.

\item[3.)] Search for single production of $H_2/A\to b\bar b$ in gluon fusion
  and possibly in association with $b\bar b$. If $H_2$ or $A$ are light, in
  the neighborhood of $200$--$250\,\gev$, the decay to $\tau^+ \tau^-$ can be
  important. It is then also possible that the heavier of the two neutral
  scalars decays to the lighter one plus a $Z$-boson.

\item[4.)] If possible, search for gluon fusion of $H_2/A\to t\bar t$
  nearer to the $\bar tt$ threshold than was done in
  Ref.~\cite{Aaboud:2017hnm}.

\item[5.)] Search for diboson resonances decaying to $V_L V_L$ and $V_L H$,
  as discussed in Sec.~2. The mass of such resonances is dictated by the
  underlying dynamics that produce the scale-invariant potential $V_0$ in
  Eq.~(\ref{eq:VzeroLP}), dynamics whose energy scale is not specified in the
  model.

\item[6.)] Drell-Yan production of $H^\pm A$, $H^\pm H_2$, $H_2 A$ and
  $H^+ H^-$ are at most a few femtobarns and may, therefore, be more
  difficult targets than $gg \to H_2,A$. On the other hand, these cross
  sections have no $\tan^2\beta$ suppression.

\item[7.)] Gluon fusion of $H_2/A\to\gamma\gamma$ may be too small to be
  detected because of the $\tan^2\beta$ suppression. If $M_{H_2/A} < 2m_t$,
  the scalar's dominant mode may be to $\bar bb$. Then
  $\sigma(gg \to H_2/A)B(H_2/A \to \gamma\gamma) \propto \tan^2\beta$, not
  $\tan^4\beta$, so there is some hope.

\item[8.)] The alignment of the 125~GeV Higgs strongly suppresses the decays
  of $H_2$ and $A$ to $W^+ W^-$ and $ZZ$, as well as $WW$ and
  $ZZ$ fusion of~$H_2$ and~$A$, providing a negative test of the model.

\item[9.)] The decay rate for $H_2 = \lim_{\delta \to 0} H_2 \to HH$ is
  suppressed by $\delta^2$, providing another negative test of the model. If
  this mode is seen, it is inconsistent with the type of model considered
  here.

\end{itemize}

\section*{4. Conclusion}

In conclusion, we have emphasized here that the low mass and apparent
Standard-Model couplings to gauge bosons and fermions of $H(125)$ can have
the {\ul{same}} symmetry origin: it is the pseudo-Goldstone boson of broken
scale symmetry, the scalon of Gildener and Weinberg~\cite{Gildener:1976ih},
and this stabilizes its mass and its alignment. In the absence of any other
example, we conjectured that the GW mechanism is the only way to achieve a
truly Higgslike dilaton. We believe this is an important theoretical
point. But there is also an important experimental one to make here. The
Gildener-Weinberg scalon picture identifies a specific mass range for new,
non-SM Higgs bosons, and that mass range is not far above
$H(125)$. Therefore, at the LHC, the relatively low region below about
$550\,\gev$ currently deserves as much attention as has been given to pushing
the machine and the detectors to their limits.

\section*{Acknowledgments}

We are grateful for conversations with and comments from several members of
the CMS and ATLAS Collaborations. We thank Erick Weinberg, Liam Fitzpatrick,
Ben Grinstein, Estia Eichten, Howard Georgi, Chris Hill, Adam Martin, Luke
Pritchett, David Sperka and Indara Suarez for valuable comments and
criticisms. KL also thanks the CERN~TH Division and, especially, Luis
\'Alvarez-Gaum\'e and Cinzia Da Via for their warm and generous hospitality
during the initial stages of this work. The work of WS was supported by the
Alexander von Humboldt Foundation in the framework of the Sofja Kovalevskaja
Award 2016, endowed by the German Federal Ministry of Education and Research.

\vfil\eject

\section*{Appendix: CP-even masses and comparison with
 Lee \& Pilaftsis} 

\begin{figure}[ht!]
 \begin{center}
\includegraphics[width=2.65in, height=2.50in]{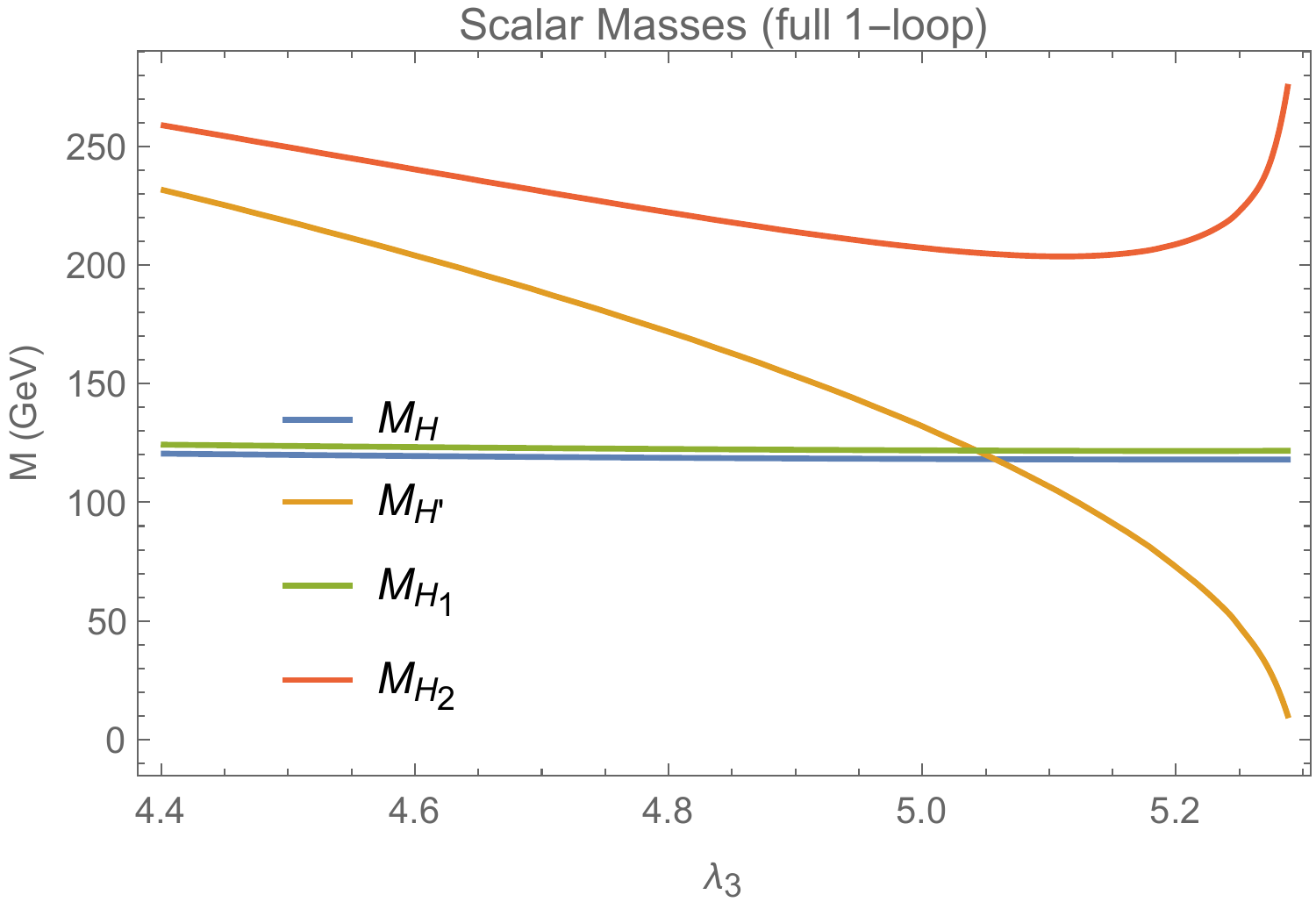}
\includegraphics[width=2.65in, height=2.50in]{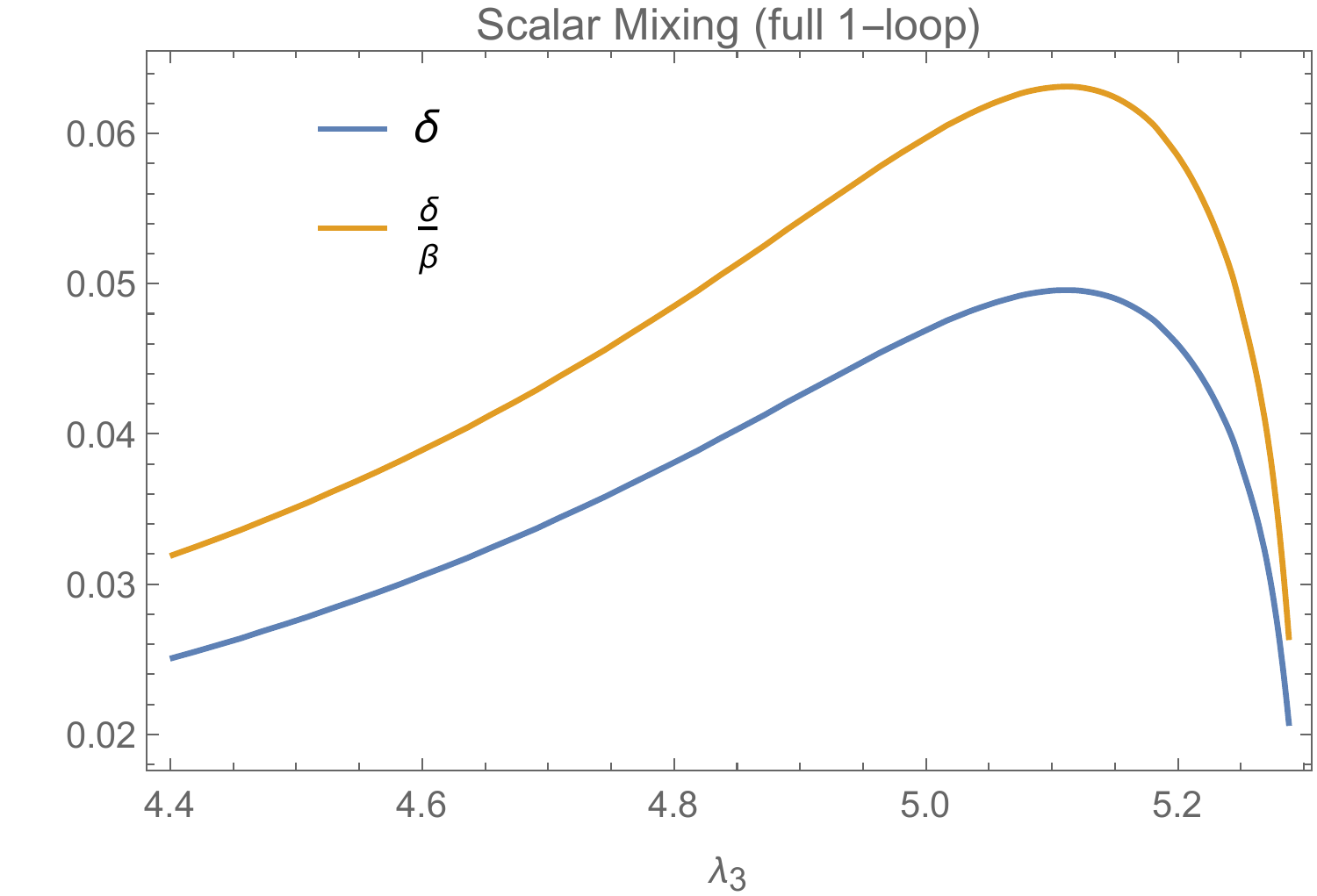}
\caption{Left: The CP-even Higgs one-loop mass eigenvalues $M_{H_1}$ and
  $M_{H_2}$, the tree-level mass $M_{H'} = \sqrt{-\lambda_{345}}\,v$ and the
  one-loop mass $M_H$ from Eq.~(\ref{eq:MHsq}) as functions of
  $\lambda_3 = (2M_{H^\pm}^2 - M_{H'}^2)/v^2$. Here, $\tan\beta = 1$ and
  $M_{H^\pm} = M_A = 400\,\gev$ corresponding to
  $\lambda_4 = \lambda_5 = -2.644$. The input $H \cong H_1$ mass is
  $M_H = 120.5\,\gev$, the corresponding initial $M_{H'} = 231.5\,\gev$ and
  $\lambda_3 = 4.403$. $M_{H'}$ vanishes at
  $\lambda_3 = 2M^2_{H^\pm}/v^2 = 5.288$. Right: The angle
  $\delta = \beta - \beta'$ and ratio $\delta/\beta$ for $\beta = \pi/4$.}
  \label{fig:oneloop}
 \end{center}
 \end{figure}
\begin{figure}[h!]
 \begin{center}
\includegraphics[width=4.00in, height=3.11in]{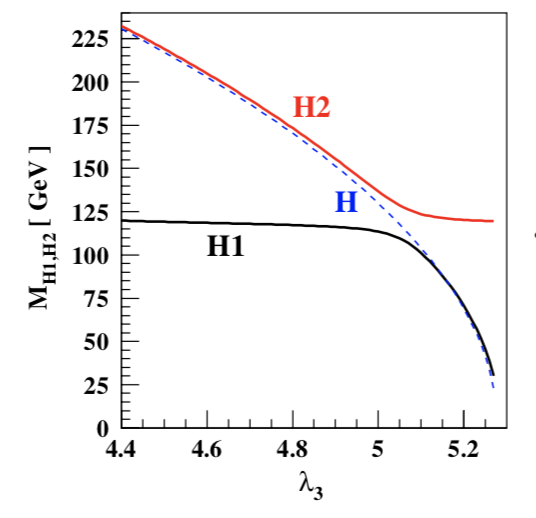}
\caption{The CP-even Higgs masses $M_{H_1},\, M_{H_2}$ as functions of
  $\lambda_3 = (2M_{H^\pm}^2 - M_{H'}^2)/v^2$ for
  $M_{H^\pm} = M_A = 400\,\gev$ and $\tan = 1$. The input $H \cong H_1$ mass
  at $\lambda_3 = 4.40$ is $M_H = 120.5\,\gev$. The dashed blue line is the
  tree-level $M_{H'} = \sqrt{-\lambda_{345}}\,v$ with
  $\lambda_4 = \lambda_5 = -M^2_A/v^2 = -2.644$. From Ref.~\cite{Lee:2012jn}.}
  \label{fig:LPfig}
 \end{center}
 \end{figure}

 We diagonalized $\CM^2_{H_{0^+}}$ with elements in Eqs.~(\ref{eq:Mpsq2}) for
 a range of $\tan\beta \le 1$ and $M_{H^\pm} = M_A \simeq 400\,\gev$. The
 general features of our results are fairly insensitive to these choices.
 The input parameters for the calculation reported here were chosen to be the
 same as those in Ref.~\cite{Lee:2012jn}, namely, $\tan\beta = 1.0$ and
 $M_{H^\pm} = M_A = 400\,\gev$. These masses determine
 $\lambda_4 = \lambda_5 = -M_A^2/v^2 = -2.644$. To initiate the calculation,
 the value $(M_H)_i = 120.5\,\gev$ was chosen from which, using
 Eq.~(\ref{eq:MHsq}), $(M_{H'})_i = 231.5\,\gev$ and $(\lambda_3)_i = 4.403$
 were determined; $\lambda_3$ was then incremented to the maximum value
 $(\lambda_3)_f = 2M^2_{H^\pm}/v^2 = 5.288$ at which
 $M^2_{H'} = -\lambda_{345}v^2$ vanishes. For each value of~$\lambda_3$, a
 new value of $M_{H'} = \sqrt{-\lambda_{345}}\,v$ is determined and used in
 the sum rule to update $M_H$ in the matrix elements of
 Eqs.~(\ref{eq:Mpsq2}). Note that it is consistent loop-perturbation theory
 to use tree-level expressions to compute the nonzero, one-loop value of
 $M_H$.

 Fig.~\ref{fig:oneloop} shows $M_H$ from Eq.~(\ref{eq:MHsq}), the
 zeroth-order mass $M_{H'}$, and the eigenvalues $M_{H_1,H_2}$ (left) and the
 angle $\delta = \beta - \beta'$ and ratio $\delta/\beta$ (right). In the
 masses plot, $M_{H_1}/M_H \cong 1.03$ for all $\lambda_3$; $M_{H_2}$ starts
 off about 10\% greater than $M_{H'}$ and increases to 70\% greater when
 $M_{H'} = M_H$ at $\lambda_3 \cong 5.04$. Then $M_{H_2}$ diverges upward
 while $M_{H'}$ plunges to zero. The mixing angle $\delta$ (right), which
 measures the deviation from perfect alignment of $H_1$, is just several
 percent and a small fraction of~$\beta$; $\delta/\beta$ has a broad maximum
 of about 6\% near $\lambda_3 = 5.11$.  For this choice of input parameters,
 then, the alignment of the $125\,\gev$ Higgs boson~$H$ is nearly
 perfect.\footnote{An extreme example takes $M_{H^\pm} = M_A = 300\,\gev$.
   Then $(M_{H'})_i = 485\,\gev$ and $(\lambda_3)_i= -0.91$. The Higgs mass
   $M_H$ calculated from the sum rule and $M_{H_1}$ remain very close as do
   $M_{H'}$ and $M_{H_2}$, and the angle $\delta =\CO(1\%)$ until near
   $(\lambda_3)_f = 2.97$ where it rises rapidly, but only to 10\%. Our
   calculations show that $\delta/\beta$ is always a few percent for all
   $\beta > 0$.}

\begin{figure}[ht!]
 \begin{center}
\includegraphics[width=2.65in, height=2.50in]{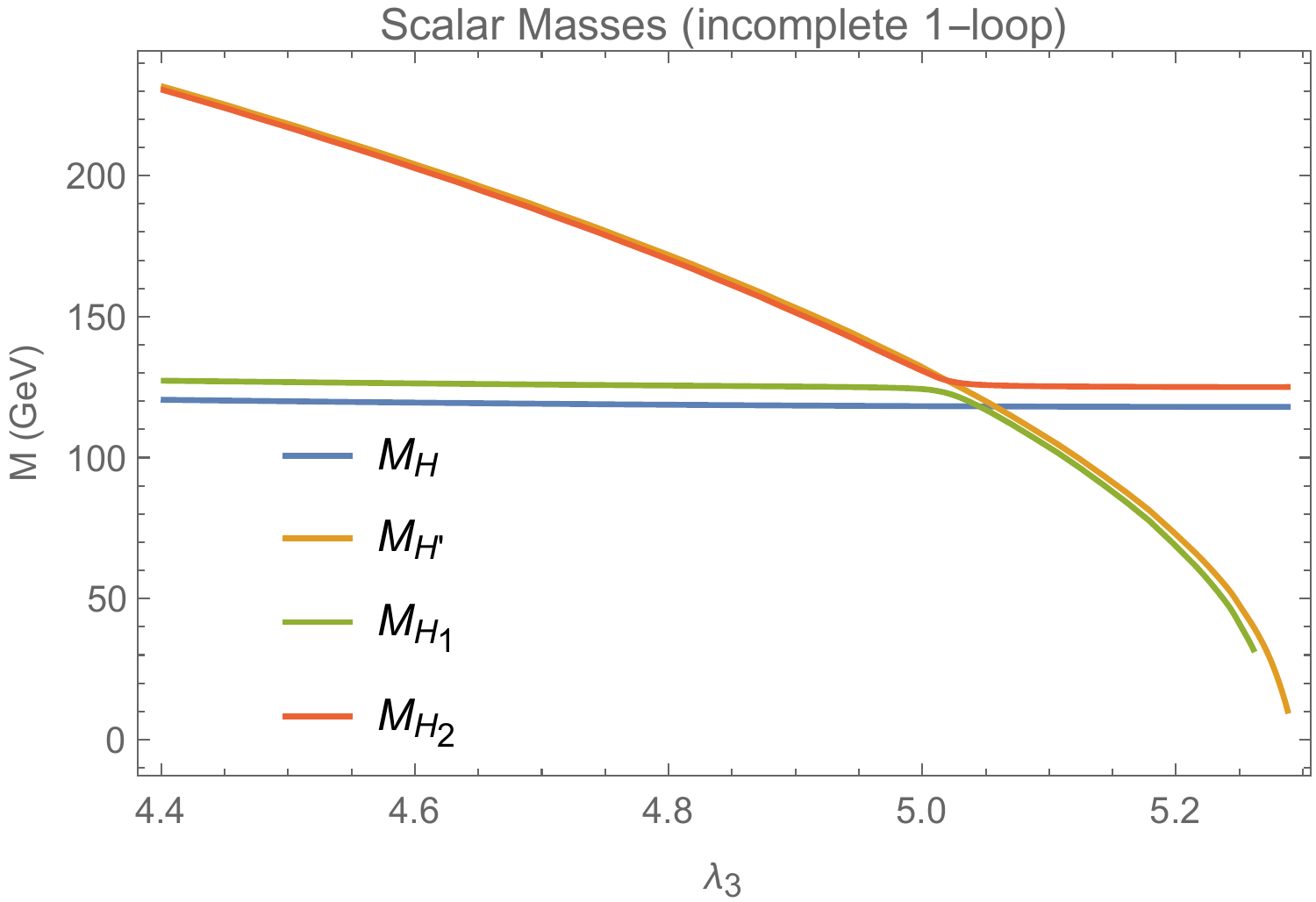}
\includegraphics[width=2.65in, height=2.50in]{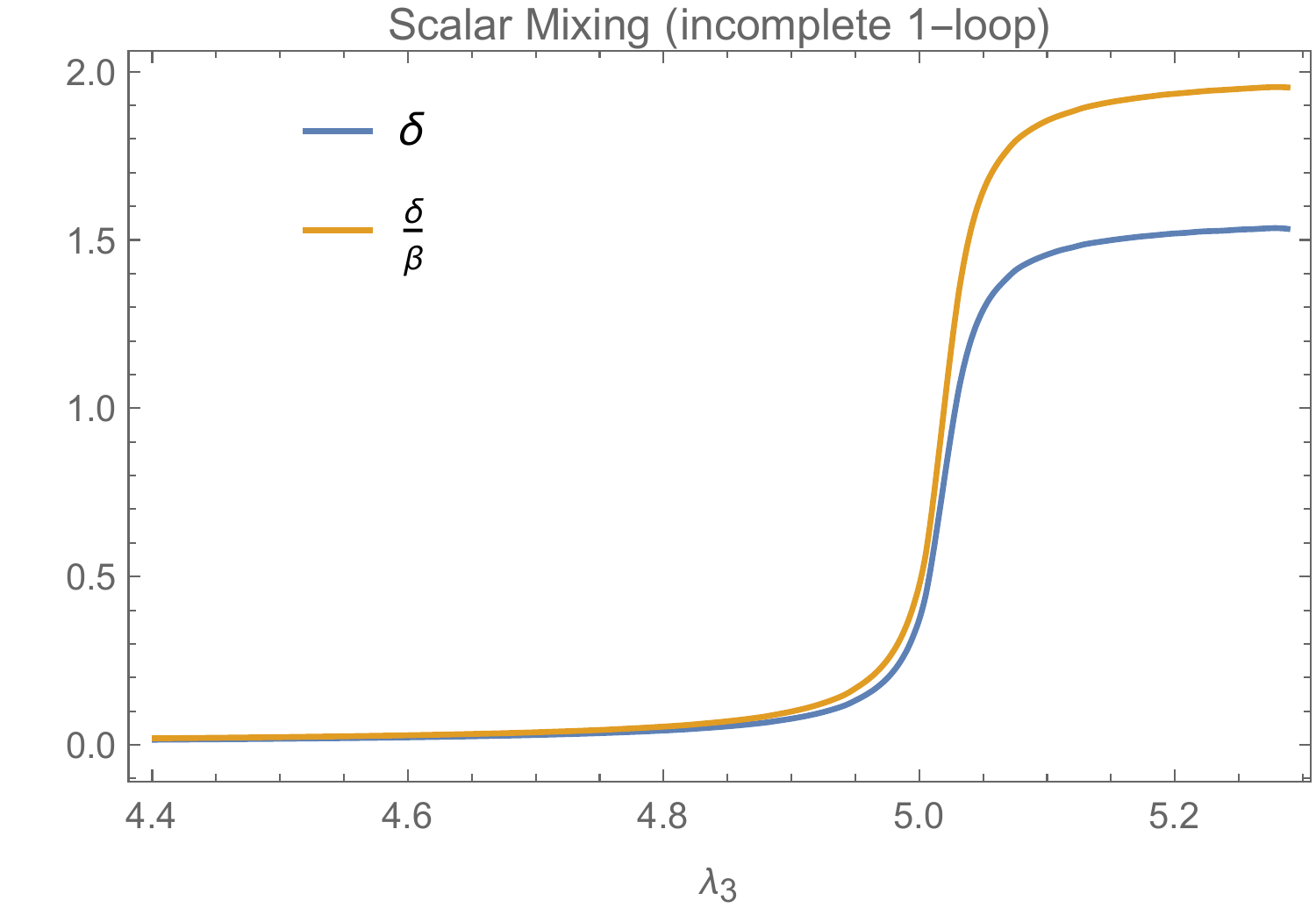}
\caption{Left: The CP-even Higgs masses, with input parameters of
  Fig~\ref{fig:oneloop}, except that the eigenvalues $M^2_{H_1,\,H_2}$ of
  $M^2_{H^{0^+}}$ were calculated using the tree-level extremal conditions
  $2\lambda_1 + \lambda_{345}\tan^2\beta = 2\lambda_2 +
  \lambda_{345} \cot^2\beta = 0$.
  Right: The angle $\delta = \beta - \beta'$ (solid) and ratio $\delta/\beta$
  (dashed) for $\beta = \pi/4$.}
  \label{fig:zeroloop}
 \end{center}
 \end{figure}

 These results are qualitatively similar to those obtained by LP in
 Ref.~\cite{Lee:2012jn}, but only up to $\lambda_3 \simeq 4.8$; see
 Fig.~\ref{fig:LPfig}. The LP paper was submitted in April 2012, before the
 announcement of the discovery of $H(125)$ and before a more precise value of
 its mass had been announced. Hence, it appears, their chosen input value of
 $M_H = 120.5\,\gev$. Up to $\lambda_3 \simeq 4.9$, $H_2 \cong H'$ with the
 tree-level mass
 $M_{H_2} \cong M_{H'} = \sqrt{-\lambda_{345}}\,v = \sqrt{2M^2_{H^\pm} -
   \lambda_3}\,v$.
 Meanwhile, $H_1 \cong H$ with $M_{H_1}$ given by Eq.~(\ref{eq:MHsq}) is
 almost constant at $120\,\gev$. In this region, $\delta$ is small and
 $\beta' \cong \beta = \pi/4$. Beyond $\lambda_3 \simeq~4.9$, there is a
 clear deviation from this behavior and a level crossing which LP identify as
 occurring at $\lambda_3^c \cong 5.06$. Above $\lambda_3^c$,
 $M_{H_2} \cong 120\,\gev$ while $M_{H_1}$ and $M_{H'}$ coalesce and fall to
 zero at $(\lambda_3)_f = 5.288$. Here, $\delta \simeq \pi/2 = 2\beta$, and
 the LP calculation is well past the point of reliable first-order
 perturbation theory.

 We cannot reproduce the level crossing seen in Fig.~\ref{fig:LPfig} using
 the $\CM^2_{H_{0^+}}$ matrix elements in Eq.~(\ref{eq:Mpsq2}). However, we
 found that we could by using the tree-level extremal conditions,
 $\lambda_1 + \thalf\lambda_{345}\tan^2\beta = \lambda_2 +
 \thalf\lambda_{345} \cot^2\beta = 0$.
 The result is illustrated in Fig.~\ref{fig:zeroloop}. The level crossing in
 the $\CM_{H_{0^+}}$ eigenvalues occurs at the same place as in LP's
 calculation. Because it is much more rapid in our calculation than in LP's,
 we can pinpoint it at $\lambda_3 = 5.04$. We do not know if this is why LP
 obtained their level crossing. But there is no doubt that using the
 tree-level extremal conditions in $\CM^2_{H_{0^+}}$ is not consistent
 loop-perturbation theory and, in fact, the results are renormalization-scale
 dependent.

\bibliography{LCH_Dilaton}

\providecommand{\href}[2]{#2}\begingroup\raggedright\begin{thebibliography}{10}

\bibitem{Aad:2012tfa}
{\bf ATLAS} Collaboration, G.~Aad {\em et.~al.}, ``{Observation of a new
  particle in the search for the Standard Model Higgs boson with the ATLAS
  detector at the LHC},'' {\em Phys.Lett.} {\bf B716} (2012) 1--29,
  \href{http://xxx.lanl.gov/abs/1207.7214}{ 1207.7214}.

\bibitem{Chatrchyan:2012ufa}
{\bf CMS} Collaboration, S.~Chatrchyan {\em et.~al.}, ``{Observation of a new
  boson at a mass of 125 GeV with the CMS experiment at the LHC},'' {\em
  Phys.Lett.} {\bf B716} (2012) 30--61,
  \href{http://xxx.lanl.gov/abs/1207.7235}{ 1207.7235}.

\bibitem{Tanabashi:2018oca}
{\bf Particle Data Group} Collaboration, M.~Tanabashi {\em et.~al.}, ``{Review
  of Particle Physics},'' {\em Phys. Rev.} {\bf D98} (2018), no.~3, 030001.

\bibitem{ATLAS:2018doi}
{\bf ATLAS} Collaboration, T.~A. collaboration, ``{Combined measurements of
  Higgs boson production and decay using up to 80 fb$^{-1}$ of proton--proton
  collision data at $\sqrt{s}=$ 13 TeV collected with the ATLAS experiment},''.

\bibitem{Sirunyan:2018hoz}
{\bf CMS} Collaboration, A.~M. Sirunyan {\em et.~al.}, ``{Observation of
  $\mathrm{t\overline{t}}$H production},'' {\em Phys. Rev. Lett.} {\bf 120}
  (2018), no.~23, 231801, \href{http://xxx.lanl.gov/abs/1804.02610}{
  1804.02610}.

\bibitem{Aaboud:2018urx}
{\bf ATLAS} Collaboration, M.~Aaboud {\em et.~al.}, ``{Observation of Higgs
  boson production in association with a top quark pair at the LHC with the
  ATLAS detector},'' \href{http://xxx.lanl.gov/abs/1806.00425}{ 1806.00425}.

\bibitem{Gunion:2002zf}
J.~F. Gunion and H.~E. Haber, ``{The CP conserving two Higgs doublet model: The
  Approach to the decoupling limit},'' {\em Phys. Rev.} {\bf D67} (2003)
  075019, \href{http://xxx.lanl.gov/abs/hep-ph/0207010}{ hep-ph/0207010}.

\bibitem{Carena:2013ooa}
M.~Carena, I.~Low, N.~R. Shah, and C.~E.~M. Wagner, ``{Impersonating the
  Standard Model Higgs Boson: Alignment without Decoupling},'' {\em JHEP} {\bf
  04} (2014) 015, \href{http://xxx.lanl.gov/abs/1310.2248}{ 1310.2248}.

\bibitem{Dev:2014yca}
P.~S. Bhupal~Dev and A.~Pilaftsis, ``{Maximally Symmetric Two Higgs Doublet
  Model with Natural Standard Model Alignment},'' {\em JHEP} {\bf 12} (2014)
  024, \href{http://xxx.lanl.gov/abs/1408.3405}{ 1408.3405}. [Erratum:
  JHEP11,147(2015)].

\bibitem{Dev:2015bta}
P.~S.~B. Dev and A.~Pilaftsis, ``{Natural Standard Model Alignment in the Two
  Higgs Doublet Model},'' {\em J. Phys. Conf. Ser.} {\bf 631} (2015), no.~1,
  012030, \href{http://xxx.lanl.gov/abs/1503.09140}{ 1503.09140}.

\bibitem{Benakli:2018vqz}
K.~Benakli, M.~D. Goodsell, and S.~L. Williamson, ``{Higgs alignment from
  extended supersymmetry},'' {\em Eur. Phys. J.} {\bf C78} (2018), no.~8, 658,
  \href{http://xxx.lanl.gov/abs/1801.08849}{ 1801.08849}.

\bibitem{Benakli:2018ldd}
K.~Benakli, Y.~Chen, and G.~Lafforgue-Marmet, ``{Predicting Alignment in a Two
  Higgs Doublet Model},''
\newblock 2018.
\newblock \href{http://xxx.lanl.gov/abs/1812.02208}{ 1812.02208}.

\bibitem{Coleman:1973jx}
S.~R. Coleman and E.~J. Weinberg, ``{Radiative Corrections as the Origin of
  Spontaneous Symmetry Breaking},'' {\em Phys. Rev.} {\bf D7} (1973)
  1888--1910.

\bibitem{Gildener:1976ih}
E.~Gildener and S.~Weinberg, ``{Symmetry Breaking and Scalar Bosons},'' {\em
  Phys. Rev.} {\bf D13} (1976) 3333.

\bibitem{Weinberg:1973ua}
S.~Weinberg, ``{Perturbative Calculations of Symmetry Breaking},'' {\em Phys.
  Rev.} {\bf D7} (1973) 2887--2910.

\bibitem{Bardeen:1995kv}
W.~A. Bardeen, ``{On naturalness in the standard model},'' in {\em {Ontake
  Summer Institute on Particle Physics Ontake Mountain, Japan, August
  27-September 2, 1995}}.
\newblock 1995.

\bibitem{Lee:2012jn}
J.~S. Lee and A.~Pilaftsis, ``{Radiative Corrections to Scalar Masses and
  Mixing in a Scale Invariant Two Higgs Doublet Model},'' {\em Phys. Rev.} {\bf
  D86} (2012) 035004, \href{http://xxx.lanl.gov/abs/1201.4891}{ 1201.4891}.

\bibitem{Hashino:2015nxa}
K.~Hashino, S.~Kanemura, and Y.~Orikasa, ``{Discriminative phenomenological
  features of scale invariant models for electroweak symmetry breaking},'' {\em
  Phys. Lett.} {\bf B752} (2016) 217--220,
  \href{http://xxx.lanl.gov/abs/1508.03245}{ 1508.03245}.

\bibitem{Goldberger:2008zz}
W.~D. Goldberger, B.~Grinstein, and W.~Skiba, ``{Distinguishing the Higgs boson
  from the dilaton at the Large Hadron Collider},'' {\em Phys.Rev.Lett.} {\bf
  100} (2008) 111802, \href{http://xxx.lanl.gov/abs/0708.1463}{ 0708.1463}.

\bibitem{Bellazzini:2012vz}
B.~Bellazzini, C.~Csaki, J.~Hubisz, J.~Serra, and J.~Terning, ``{A Higgslike
  Dilaton},'' {\em Eur. Phys. J.} {\bf C73} (2013), no.~2, 2333,
  \href{http://xxx.lanl.gov/abs/1209.3299}{ 1209.3299}.

\bibitem{Serra:2013kga}
J.~Serra, ``{A higgs-like dilaton: viability and implications},'' {\em EPJ Web
  Conf.} {\bf 60} (2013) 17005, \href{http://xxx.lanl.gov/abs/1312.0259}{
  1312.0259}.

\bibitem{Bellazzini:2014yua}
B.~Bellazzini, C.~Csáki, and J.~Serra, ``{Composite Higgses},'' {\em
  Eur.Phys.J.} {\bf C74} (2014), no.~5, 2766,
  \href{http://xxx.lanl.gov/abs/1401.2457}{ 1401.2457}.

\bibitem{Ghilencea:2016dsl}
D.~M. Ghilencea, Z.~Lalak, and P.~Olszewski, ``{Standard Model with
  spontaneously broken quantum scale invariance},'' {\em Phys. Rev.} {\bf D96}
  (2017), no.~5, 055034, \href{http://xxx.lanl.gov/abs/1612.09120}{
  1612.09120}.

\bibitem{Hernandez-Leon:2017kea}
P.~Hernandez-Leon and L.~Merlo, ``{Distinguishing A Higgs-Like Dilaton Scenario
  With A Complete Bosonic Effective Field Theory Basis},'' {\em Phys. Rev.}
  {\bf D96} (2017), no.~7, 075008, \href{http://xxx.lanl.gov/abs/1703.02064}{
  1703.02064}.

\bibitem{Bellazzini:2013fga}
B.~Bellazzini, C.~Csaki, J.~Hubisz, J.~Serra, and J.~Terning, ``{A Naturally
  Light Dilaton and a Small Cosmological Constant},'' {\em Eur. Phys. J.} {\bf
  C74} (2014) 2790, \href{http://xxx.lanl.gov/abs/1305.3919}{ 1305.3919}.

\bibitem{Coradeschi:2013gda}
F.~Coradeschi, P.~Lodone, D.~Pappadopulo, R.~Rattazzi, and L.~Vitale, ``{A
  naturally light dilaton},'' {\em JHEP} {\bf 11} (2013) 057,
  \href{http://xxx.lanl.gov/abs/1306.4601}{ 1306.4601}.

\bibitem{Glashow:1976nt}
S.~L. Glashow and S.~Weinberg, ``{Natural Conservation Laws for Neutral
  Currents},'' {\em Phys. Rev.} {\bf D15} (1977) 1958.

\bibitem{Lane:2015fza}
K.~Lane and L.~Pritchett, ``{Heavy Vector Partners of the Light Composite
  Higgs},'' {\em Phys. Lett.} {\bf B753} (2016) 211--214,
  \href{http://xxx.lanl.gov/abs/1507.07102}{ 1507.07102}.

\bibitem{Appelquist:2015vdl}
T.~Appelquist, Y.~Bai, J.~Ingoldby, and M.~Piai, ``{Spectrum-doubled Heavy
  Vector Bosons at the LHC},'' {\em JHEP} {\bf 01} (2016) 109,
  \href{http://xxx.lanl.gov/abs/1511.05473}{ 1511.05473}.

\bibitem{Brooijmans:2016vro}
G.~Brooijmans {\em et.~al.}, ``{Les Houches 2015: Physics at TeV colliders -
  new physics working group report},'' in {\em {9th Les Houches Workshop on
  Physics at TeV Colliders (PhysTeV 2015) Les Houches, France, June 1-19,
  2015}}.
\newblock 2016.
\newblock \href{http://xxx.lanl.gov/abs/1605.02684}{ 1605.02684}.

\bibitem{Cavaliere:2018zcf}
V.~Cavaliere, R.~Les, T.~Nitta, and K.~Terashi, ``{HE-LHC prospects for diboson
  resonance searches and electroweak WW/WZ production via vector boson
  scattering in the semi-leptonic final states},''
  \href{http://xxx.lanl.gov/abs/1812.00841}{ 1812.00841}.

\bibitem{Carena:2002es}
M.~Carena and H.~E. Haber, ``{Higgs boson theory and phenomenology},'' {\em
  Prog. Part. Nucl. Phys.} {\bf 50} (2003) 63--152,
  \href{http://xxx.lanl.gov/abs/hep-ph/0208209}{ hep-ph/0208209}.

\bibitem{Hill:2014mqa}
C.~T. Hill, ``{Is the Higgs Boson Associated with Coleman-Weinberg Dynamical
  Symmetry Breaking?},'' {\em Phys. Rev.} {\bf D89} (2014), no.~7, 073003,
  \href{http://xxx.lanl.gov/abs/1401.4185}{ 1401.4185}.

\bibitem{Misiak:2006zs}
M.~Misiak {\em et.~al.}, ``{Estimate of $\mathcal{B} (\bar B \to X_s \gamma)$
  at $O(\alpha_s^2)$},'' {\em Phys. Rev. Lett.} {\bf 98} (2007) 022002,
  \href{http://xxx.lanl.gov/abs/hep-ph/0609232}{ hep-ph/0609232}.

\bibitem{Branco:2011iw}
G.~C. Branco, P.~M. Ferreira, L.~Lavoura, M.~N. Rebelo, M.~Sher, and J.~P.
  Silva, ``{Theory and phenomenology of two-Higgs-doublet models},'' {\em Phys.
  Rept.} {\bf 516} (2012) 1--102, \href{http://xxx.lanl.gov/abs/1106.0034}{
  1106.0034}.

\bibitem{Kanemura:1993hm}
S.~Kanemura, T.~Kubota, and E.~Takasugi, ``{Lee-Quigg-Thacker bounds for Higgs
  boson masses in a two doublet model},'' {\em Phys. Lett.} {\bf B313} (1993)
  155--160, \href{http://xxx.lanl.gov/abs/hep-ph/9303263}{ hep-ph/9303263}.

\bibitem{Kennedy:1988sn}
D.~C. Kennedy and B.~W. Lynn, ``{Electroweak Radiative Corrections with an
  Effective Lagrangian: Four Fermion Processes},'' {\em Nucl. Phys.} {\bf B322}
  (1989) 1.

\bibitem{Peskin:1990zt}
M.~E. Peskin and T.~Takeuchi, ``A new constraint on a strongly interacting
  Higgs sector,'' {\em Phys. Rev. Lett.} {\bf 65} (1990) 964--967.

\bibitem{Peskin:1991sw}
M.~E. Peskin and T.~Takeuchi, ``{Estimation of oblique electroweak
  corrections},'' {\em Phys. Rev.} {\bf D46} (1992) 381--409.

\bibitem{Golden:1990ig}
M.~Golden and L.~Randall, ``Radiative corrections to electroweak parameters in
  technicolor theories,'' {\em Nucl. Phys.} {\bf B361} (1991) 3--23.

\bibitem{Holdom:1990tc}
B.~Holdom and J.~Terning, ``Large corrections to electroweak parameters in
  technicolor theories,'' {\em Phys. Lett.} {\bf B247} (1990) 88--92.

\bibitem{Altarelli:1991fk}
G.~Altarelli, R.~Barbieri, and S.~Jadach, ``Toward a model independent analysis
  of electroweak data,'' {\em Nucl. Phys.} {\bf B369} (1992) 3--32.

\bibitem{Battye:2011jj}
R.~A. Battye, G.~D. Brawn, and A.~Pilaftsis, ``{Vacuum Topology of the Two
  Higgs Doublet Model},'' {\em JHEP} {\bf 08} (2011) 020,
  \href{http://xxx.lanl.gov/abs/1106.3482}{ 1106.3482}.

\bibitem{Pilaftsis:2011ed}
A.~Pilaftsis, ``{On the Classification of Accidental Symmetries of the Two
  Higgs Doublet Model Potential},'' {\em Phys. Lett.} {\bf B706} (2012)
  465--469, \href{http://xxx.lanl.gov/abs/1109.3787}{ 1109.3787}.

\bibitem{Patrignani:2016xqp}
{\bf Particle Data Group} Collaboration, C.~Patrignani {\em et.~al.}, ``{Review
  of Particle Physics},'' {\em Chin. Phys.} {\bf C40} (2016), no.~10, 100001.

\bibitem{Toussaint:1978zm}
D.~Toussaint, ``{Renormalization Effects From Superheavy Higgs Particles},''
  {\em Phys. Rev.} {\bf D18} (1978) 1626.

\bibitem{Kanemura:2011sj}
S.~Kanemura, Y.~Okada, H.~Taniguchi, and K.~Tsumura, ``{Indirect bounds on
  heavy scalar masses of the two-Higgs-doublet model in light of recent Higgs
  boson searches},'' {\em Phys. Lett.} {\bf B704} (2011) 303--307,
  \href{http://xxx.lanl.gov/abs/1108.3297}{ 1108.3297}.

\bibitem{Abbiendi:2013hk}
{\bf LEP, DELPHI, OPAL, ALEPH, L3} Collaboration, G.~Abbiendi {\em et.~al.},
  ``{Search for Charged Higgs bosons: Combined Results Using LEP Data},'' {\em
  Eur. Phys. J.} {\bf C73} (2013) 2463,
  \href{http://xxx.lanl.gov/abs/1301.6065}{ 1301.6065}.

\bibitem{Khachatryan:2015qxa}
{\bf CMS} Collaboration, V.~Khachatryan {\em et.~al.}, ``{Search for a charged
  Higgs boson in pp collisions at $ \sqrt{s}=8 $ TeV},'' {\em JHEP} {\bf 11}
  (2015) 018, \href{http://xxx.lanl.gov/abs/1508.07774}{ 1508.07774}.

\bibitem{Aaboud:2017hnm}
{\bf ATLAS} Collaboration, M.~Aaboud {\em et.~al.}, ``{Search for Heavy Higgs
  Bosons $A/H$ Decaying to a Top Quark Pair in $pp$ Collisions at
  $\sqrt{s}=8\text{ }\text{ }\mathrm{TeV}$ with the ATLAS Detector},'' {\em
  Phys. Rev. Lett.} {\bf 119} (2017), no.~19, 191803,
  \href{http://xxx.lanl.gov/abs/1707.06025}{ 1707.06025}.

\bibitem{Aaboud:2017sjh}
{\bf ATLAS} Collaboration, M.~Aaboud {\em et.~al.}, ``{Search for additional
  heavy neutral Higgs and gauge bosons in the ditau final state produced in 36
  fb$^{−1}$ of pp collisions at $ \sqrt{s}=13 $ TeV with the ATLAS
  detector},'' {\em JHEP} {\bf 01} (2018) 055,
  \href{http://xxx.lanl.gov/abs/1709.07242}{ 1709.07242}.

\bibitem{Sirunyan:2018taj}
{\bf CMS} Collaboration, A.~M. Sirunyan {\em et.~al.}, ``{Search for beyond the
  standard model Higgs bosons decaying into a $\mathrm{b\overline{b}}$ pair in
  pp collisions at $\sqrt{s} =$ 13 TeV},''
  \href{http://xxx.lanl.gov/abs/1805.12191}{ 1805.12191}.

\bibitem{Khachatryan:2016are}
{\bf CMS} Collaboration, V.~Khachatryan {\em et.~al.}, ``{Search for neutral
  resonances decaying into a Z boson and a pair of b jets or $\tau$ leptons},''
  {\em Phys. Lett.} {\bf B759} (2016) 369--394,
  \href{http://xxx.lanl.gov/abs/1603.02991}{ 1603.02991}.

\bibitem{Aaboud:2018eoy}
{\bf ATLAS} Collaboration, M.~Aaboud {\em et.~al.}, ``{Search for a heavy Higgs
  boson decaying into a $Z$ boson and another heavy Higgs boson in the
  $\ell\ell bb$ final state in $pp$ collisions at $\sqrt{s}=13$ TeV with the
  ATLAS detector},'' {\em Phys. Lett.} {\bf B783} (2018) 392--414,
  \href{http://xxx.lanl.gov/abs/1804.01126}{ 1804.01126}.

\bibitem{Aaboud:2018cwk}
{\bf ATLAS} Collaboration, M.~Aaboud {\em et.~al.}, ``{Search for charged Higgs
  bosons decaying into top and bottom quarks at $\sqrt{s}$ = 13 TeV with the
  ATLAS detector},'' \href{http://xxx.lanl.gov/abs/1808.03599}{ 1808.03599}.

\end{thebibliography}\endgroup
\bibliographystyle{utcaps}
\end{document}